\documentclass[iop,twocolappendix]{emulateapj}

\usepackage{epsfig}
\usepackage{epstopdf}
\usepackage{xcolor}
\usepackage{natbib}

\begin{document}

%@arxiver{X_vs_Fe_N1023.eps,N1023_N2974.eps}

\def\beq{\begin{equation}}
\def\eeq{\end{equation}}

\newcommand{\na}{\rm Na \, I}
\newcommand{\cat}{\rm Ca \, II}
\newcommand{\feh}{\rm FeH}
\newcommand{\tio}{\rm TiO}
\newcommand{\feavg}{\left<{\rm Fe}\right>}
\newcommand{\kms}{\,{\rm km\,s^{-1}}}
\newcommand{\msun}{\, M_\odot}
\newcommand{\lsun}{\, L_\odot}
\newcommand{\lvsun}{\, L_{\odot,V}}
\newcommand{\mlrsun}{\, M_{\odot} L_{\odot,R}^{-1}}
\newcommand{\mlvsun}{\, M_{\odot} L_{\odot,V}^{-1}}
\newcommand{\mbh}{M_\bullet}
\newcommand{\mstar}{M_\star}
\newcommand{\mbulge}{M_{\rm bulge}}
\newcommand{\ml}{M_\star/L}
\newcommand{\mlr}{M_\star/L_R}
\newcommand{\mlv}{M_\star/L_V}
\newcommand{\mlk}{\Upsilon_K}
\newcommand{\mlkmw}{\Upsilon_{K, \rm{MW}}}
\newcommand{\rinf}{\, r_{\rm inf}}
\newcommand{\reff}{\,r_{\rm eff}}
\newcommand{\teff}{T_{\rm eff}}
\newcommand{\mathp}{\mathcal{P}}
\newcommand{\scut}{\sigma_{\rm cut}}
\newcommand{\atlas}{{\rm ATLAS^{3D}}}
\newcommand {\gtsim} {\ {\raise-.5ex\hbox{$\buildrel>\over\sim$}}\ }
\newcommand {\ltsim} {\ {\raise-.5ex\hbox{$\buildrel<\over\sim$}}\ } 

\defcitealias{CvD12a}{CvD12}

\title{Radial Trends in IMF-Sensitive Absorption Features in Two Early-Type Galaxies: Evidence for Abundance-Driven Gradients}

\author{Nicholas J. McConnell \footnotemark[1,2], Jessica R. Lu \footnotemark[1], and Andrew W. Mann \footnotemark[3]}

\footnotetext[1]{Institute for Astronomy, University of Hawaii at M\={a}noa, Honolulu, HI; njmcconnell@gmail.com}
\footnotetext[2]{NRC Herzberg, Victoria, BC, Canada}
\footnotetext[3]{Department of Astronomy, University of Texas at Austin, Austin, TX}

\begin{abstract}

Samples of early-type galaxies show a correlation between stellar velocity dispersion and the stellar initial mass function (IMF) as inferred from gravity-sensitive absorption lines in the galaxies' central regions.  To search for spatial variations in the IMF,
we have observed two early-type galaxies with Keck/LRIS and measured radial gradients in the strengths of absorption features from 4000-5500 \AA\, and 8000-10,000 \AA.  We present spatially resolved measurements of the dwarf-sensitive spectral indices $\na$ (8190 \AA) and Wing-Ford $\feh$ (9915 \AA), as well as indices for species of H, C$_2$, CN, Mg, Ca, $\tio$, and Fe.
Our measurements show a metallicity gradient in both objects, and Mg/Fe
consistent with a shallow gradient in $\alpha$-enhancement, matching widely observed trends for massive early-type galaxies.  The $\na$ index and the CN$_1$ index at 4160 \AA\, exhibit significantly steeper gradients, with a break at $r \sim 0.1 \reff$ ($r \sim 300$ pc).  
Inside this radius $\na$ strength increases sharply toward the galaxy center, consistent with a rapid central rise in [Na/Fe].
In contrast, the ratio of $\feh$ to Fe index strength decreases toward the galaxy center.  This behavior cannot be reproduced by a steepening IMF inside $0.1 \reff$ if the IMF is a single power law.  
While gradients in the mass function above $\sim 0.4 \msun$ may occur, exceptional care is required to disentangle these IMF variations from the extreme variations in individual element abundances near the galaxies' centers.  \\

\end{abstract}

\pagestyle{plain}
\pagenumbering{arabic}
%\addtolength{\topmargin}{-0.5in}

\maketitle

\section{Introduction}
\label{sec:intro}

Driven primarily by observations within our own Galaxy, the assumption of a universal stellar initial mass function (IMF) throughout space and time has been employed in numerous studies of galaxy evolution. The canonical IMFs have the form of a power-law at high masses with d$N$/d$m$ $\propto m^{-\alpha}$ and $\alpha=2.3$ above $m > 0.5 \msun$ \citep{Kroupa01} or $m > 1 \msun$ \citep{Chabrier03}. These IMFs then flatten and turn-over with a decreasing number of stars at masses below 0.5-1 $\msun$. In an extra-galactic context, key observables are influenced by the universal-IMF assumption, such as the galaxy mass function, mass-metallicity relation, and the correlation between total star formation rate and galaxy mass 
\citep[e.g.,][]{Noeske07,Wuyts11,Leauthaud12,Smit12,Behroozi13,Zahid14,Salmon15}.  
Spectral energy density (SED) fitting is the most prevalent method for determining the stellar masses of intermediate- and high-redshift galaxies, and requires an assumed form of the IMF \citep[e.g.,][]{Conroy09,Marchesini09}.
Consequently, claims of IMF variations deserve intense scrutiny \citep{Bastian10,Krumholz14} since any systemic IMF variation has broad implications for the inferred properties of galaxies and galaxy evolution.

In the past five years numerous studies have asserted that early-type galaxies with the largest velocity dispersions ($\sigma$) reveal a bottom-heavy IMF in their old stellar populations: an overabundance of stars with $m < 1 \msun$ relative to the canonical IMFs of \citet{Kroupa01} or \citet{Chabrier03}.
Methodologies used to assess the IMF in these galaxies have included (1) examinations of stellar absorption features dominated by either giant or dwarf stars 
\citep[e.g.,][]{Cenarro03,vDC10,vDC12,CvD12b,Ferreras13,LaBarbera13,Spiniello14},
(2) comparisons of mass-to-light ratios from stellar population synthesis (SPS) and stellar dynamics
\citep[e.g.,][]{Capp12,Capp13b,Dutton13b,McDermid14},
(3) and comparisons between SPS and gravitational lensing
\citep[e.g.,][]{Treu10,Barnabe13,Posacki15,Spiniello11,Spiniello12,Spiniello15b}.
Results with the most divergent IMFs show that the inferred slope of the IMF above $0.1 \msun$ steepens to $\alpha \sim 3$ for the most massive early-type galaxies \citep[e.g.,][]{LaBarbera13}.
Other studies have found a deviation between the average IMF in early-type galaxy samples and the canonincal IMF, 
without verifying a differential trend between different early-type galaxies
\citep[e.g.,][]{Auger10,Dutton12,Smith12,Smith14}.
Alternatively, a few investigations have identified massive early-type galaxies with an IMF similar to the Milky Way
\citep[e.g.,][]{Smith13,Smith15a}. 
Using a different approach, \citet{Peacock14} examined eight nearby galaxies for emission from X-ray binaries, and found that the fraction of massive stellar remnants across their sample was consistent with a uniform IMF slope above $8 \msun$.
Thus, it is still debated whether the IMF varies, how much the functional form changes, and how these variations depend on galaxy properties.

A limitation of the majority of IMF investigations in early-type galaxies is the use of a single spatial aperture per galaxy.  For instance, stacked spectra from the Sloan Digital Sky Survey (SDSS) have a fixed radius on the sky, blending data from less than and greater than one galaxy effective radius, $\reff$, from different galaxies
\citep[e.g.,][]{Ferreras13,LaBarbera13,Spiniello14}.
In contrast, the absorption-line studies by \citet{vDC12} and \citet{CvD12b} focus on the innermost regions of nearby galaxies, with a long-slit aperture of $\reff/8$.
For lensing studies, the Einstein radius matches a different physical radius in each galaxy, typically between $\sim 0.3 \reff$ and $\sim 1 \reff$ \citep[e.g.][]{Koopmans09,Smith15a}.
\citet{Capp12,Capp13b} and \citet{McDermid14} use resolved two-dimensional stellar kinematics, extending to 1 $\reff$ for many of the galaxies in their $\atlas$ sample.  However, the most massive galaxies in $\atlas$ are not covered out to 1 $\reff$. 
At best, studies with different spatial footprints offer leverage for interpreting the role of IMF gradients within individual galaxies, although there are numerous complications from synthesizing heterogeneous and sometimes contradictory results. 
At worst, these studies all fail to distinguish between an IMF that varies only from galaxy to galaxy and IMF gradients within single galaxies.
 
Theoretical motivation for the presence of IMF gradients within early-type galaxies comes from models of inside-out growth, wherein massive galaxies are built first as compact starbursts and then accrete numerous smaller systems at large radii 
\citep[e.g.,][]{Naab09,Naab14,Hopkins10,Oser12,Shankar13}.
This model coincides with observations of size growth in massive red galaxies from redshifts $\sim 2$ to the present
\citep[e.g.,][]{Trujillo06,Dokkum10,Patel13,vanderWel14,Vulcani14}, 
and with the metal-poor stellar halos of nearby early-type galaxies 
\citep[e.g.,][]{CGA10,Pastorello14,Greene12,Greene15}.
If the IMF differs between low- and high-$\sigma$ galaxies, then the most massive (and highest $\sigma$) early-type galaxies should naturally exhibit IMF gradients, as their outer regions have been assembled from smaller systems.

A second challenge for IMF investigations in early-type galaxies is the competing influence of elemental abundance ratios, which can drastically reshape stellar absorption features and subtly alter the mass-to-light ratio of stars.  Some SPS models can vary the abundances of individual elements \citep[e.g.,][]{Graves08,CvD12a}.  Yet the impacts of these abundance variations are woefully entangled with one another, and with the effects of IMF variations and other systematics such as isochrone offsets in temperature-luminosity space
\citep{Graves08,CvD12a,Spiniello15a}.
These same models have identified stellar absorption features that are especially sensitive to IMF variations, and targeted analyses of those features in observed galaxy spectra claim sufficient leverage to detect IMF variations robustly \citep[e.g.,][]{CvD12b,LaBarbera13,Spiniello14}.  
However, abundance ratios are known to vary within individual galaxies
\citep[e.g.,][]{Strom76,Tamura00,Weijmans09,Kuntschner10,Greene12,Greene13,Greene15},
and the few investigations of IMF gradients in individual systems have offered scant analysis of single-element abundance variations \citep{MartinNavarro15,MN15b,MN15c}.

Herein we examine two early-type galaxies, NGC 1023 and NGC 2974, for IMF and abundance gradients.  For each object, we use long-slit spectra to probe spatial scales from $\sim 100$ pc to a few kpc ($0.03 \reff$ to $\sim 1 \reff$).  
We present gradients in a selected set of stellar absorption line indices and qualitatively interpret their connection to stellar population gradients.
We have paid special attention to the sodium doublet at 8190 \AA\, (hereafter $\na$) and the Wing-Ford iron hydride feature at 9915 \AA\, (hereafter $\feh$), both of which are sensitive to the number density of cool dwarf stars.  Our spectra also cover the giant-sensitive calcium triplet at 8500-8660 \AA\, (hereafter $\cat$), temperature-sensitive $\tio$ features, numerous features of individual atomic species, and several Balmer lines.  

Recently, \citet{MartinNavarro15,MN15b} analyzed long-slit spectra of four nearby early-type galaxies and reported IMF gradients in two ellipticals with large central $\sigma$ (270-300 $\kms$) and a mild gradient in the compact, high-$\sigma$ galaxy NGC 1277.  They inferred a uniform IMF slope in an elliptical galaxy with  $\sigma \approx 100 \kms$. 
The two objects presented herein have central $\sigma \approx 210$-250 $\kms$, and our set of IMF-sensitive spectral indices has little overlap with those analyzed by \citet{MartinNavarro15,MN15b}.  Although our spectral coverage excludes the $\tio_1$ and $\tio_2$ features at 5960 \AA\, and 6230 \AA, we perform much more rigorous analysis of $\na$ and are among the first to present spatially resolved measurements of $\feh$.

We summarize the basic properties of NGC 1023 and NGC 2974 in Table~\ref{tab:sample}.  NGC 1023 is an SB0 galaxy, and NGC 2974 is an E4, as classified by the NASA/IPAC Extragalactic Database.  Both galaxies are fast rotators \citep{Emsellem11}.  
We selected both objects from the sample of \citet{vDC12}, who observed the central $\reff/8$ of 34 early-type galaxies with identical spectral coverage to our investigation.  Based on full-spectrum fitting to stellar population models, \citet{CvD12b} 
measured a stellar mass-to-light ratio, $\mlk$, of $1.53 \mlkmw$ in NGC 1023, and $1.45 \mlkmw$ in NGC 2974, 
where $\mlkmw$ is the stellar mass-to-light ratio of the Milky Way disk.  They infer that the central IMF in both galaxies is consistent with a \citet{Salpeter55} power-law ($\alpha = 2.35$) extending down to $0.1 \msun$ and significantly more bottom-heavy than the Kroupa or Chabrier forms.

This paper is organized as follows.  We summarize our long-slit observations in \S\ref{sec:obs} and our data reduction methods in \S\ref{sec:data}.  In \S\ref{sec:indices} we describe how absorption line indices are determined from our spectra, including random and systematic errors.  We present radial trends for 13 selected line indices in \S\ref{sec:gradients}.  In \S\ref{sec:disc} we offer a qualitative interpretation of the underlying stellar population trends in NGC 1023 and NGC 2974, with rigorous stellar population modeling deferred for future work.  Our conclusions are briefly summarized in \S\ref{sec:conc}.  Our two appendices contain detailed examinations of systematic errors in line index measurements, with particular focus on the $\na$ feature.

%TABLE - sample properties and photometric parameters
%
%{table*} sets table across whole page in emulateapj5
\begin{table*}[t]
%\begin{small}
\begin{center}
\caption{Targets and Observations}
\label{tab:sample}
%\leavevmode
\begin{tabular}[b]{cccccccccccc}  %each c is a column, [c]entered.  For lines between columns, [c|c|c|c] etc.
\hline
Galaxy & $D$ & $M_K$ & $\reff$ & $\sigma_c$ & [Fe/H]$_c$ & [Mg/Fe]$_c$ & [Z/H]$_c$ & [$\alpha$/Fe]$_c$ & Age$_c$ &  Integration time & Slit PA \\
\smallskip
 & (Mpc) & &  ($''$) & ($\kms$) & dex & dex & dex & dex & Gyr & (s) & ($^\circ$) \\
\smallskip
(1) & (2) & (3) & (4) & (5) & (6) & (7) & (8) & (9) & (10) & (11) & (12) \\
\hline 
\\
NGC 1023 & 11.1 & -24.01 & 47.8 & 217 & -0.01 & +0.18 & +0.09 & +0.19 & 11.7 &  $13 \times 600$ & 85 \\  %'r' chip at 85 deg.
NGC 2974 & 20.9 & -23.62 & 38.0 & 247 & -0.06 & +0.20 & +0.11 & +0.29 & 8.9 & $14 \times 600$ & 45 \\   %'r' chip at 225 deg.
%NGC 4486 & 17.2 & -25.31 & 81.2 & 385 & $X \times 900$ & 152 \\  % 'r' chip at 152 deg.
\hline
\end{tabular}
\end{center}
\begin{small}
Notes: We adopt distances, $K$-band absolute magnitudes, and effective radii (columns 2-4) from the $\atlas$ survey \citep{Capp11}.  Central velocity dispersions, [Fe/H], and [Mg/Fe] (columns 5-7) are from \citet{CvD12b} and approximate a circular aperture with radius $\reff$/8.
Central [Z/H], [$\alpha$/Fe], and stellar ages (columns 8-10) are from \citet{Kuntschner10}, summing SAURON integral-field data over a circular aperture of $\reff/8$.  Using the new data herein and stellar population models from \citet{CvD12a}, we estimate that both galaxies are $> 10$ Gyr old at all radii.
\end{small}
\vspace{0.15in}
\end{table*}

\section{Observations}
\label{sec:obs}

We observed NGC 1023 and NGC 2974 in December 2013 using LRIS on the Keck 1 telescope \citep{Oke95,Rockosi10}, in long-slit mode.  
LRIS comprises a red arm and a blue arm.  For the red side, we used the 600 line mm$^{-1}$ grating spanning 7500-10,800 \AA, sampled at $\approx 0.79$ \AA\, per pixel.  On the blue side we used the 600 line mm$^{-1}$ grism spanning 3100-5560 \AA, sampled at $\approx 0.62$ \AA\, per pixel.
For this instrumental setup we measured a spectral resolution of $\approx 3.0$ \AA\, full width at half-maximum (FWHM) for both the blue arm and the red arm.
Our slit was $0.7''$ wide and spanned a length of $175''$.  The LRIS field of view is sampled by two detectors on each spectrograph arm, resulting in a coverage gap near the center of the slit.
For the data presented herein, the gap spans $35''$ on the red side and $14''$ on the blue side.  On the LRIS detectors we employed a pixel scale of $0.135''$ in the spatial direction.  Seeing on the night of observations was between $1.0''$ and $1.5''$ FWHM.  

We placed our slit along the major axis of each galaxy with the galaxy center slightly offset from the chip gap. 
For a single exposure our spatial coverage extended from $\approx 60''$ on one side of the galaxy to $\approx 115''$ on the opposite side.  
Images were dithered between two positions that straddled the gap to achieve symmetric coverage of each galaxy.  Additionally, a sky field $4'$ away was observed after every two to four science exposures. 
Total integration times for each target are included in Table~\ref{tab:sample}.  We completed more than two hours of integration time on each galaxy in order to obtain signal-to-noise ratios ($S/N$) $\sim 100$ per pixel in spatial bins near $1 \reff$.  The final spectra for NGC 1023 meet this criterion out to $1.6 \reff$, for all features except the faint $\feh$ band near 1 $\mu$m.  For NGC 2974 the outermost bin with sufficient $S/N$ spans 0.5-0.8 $\reff$.

Although our instrument setup matched that of \citet{vDC12}, our observing pattern was selected to maximize spatial coverage and measure gradients in absorption line depths.  
In contrast, observations by \citet{vDC12} aligned the slit along the minor axis of each galaxy and used the far edges for in-frame sky subtraction.  As a result, their previous analysis was restricted to the central few arc seconds of each galaxy, corresponding to an aperture of $\reff/8$.

\section{Data Processing and Analysis}
\label{sec:data}

\begin{figure*}[!h]
\centering
  \epsfig{figure=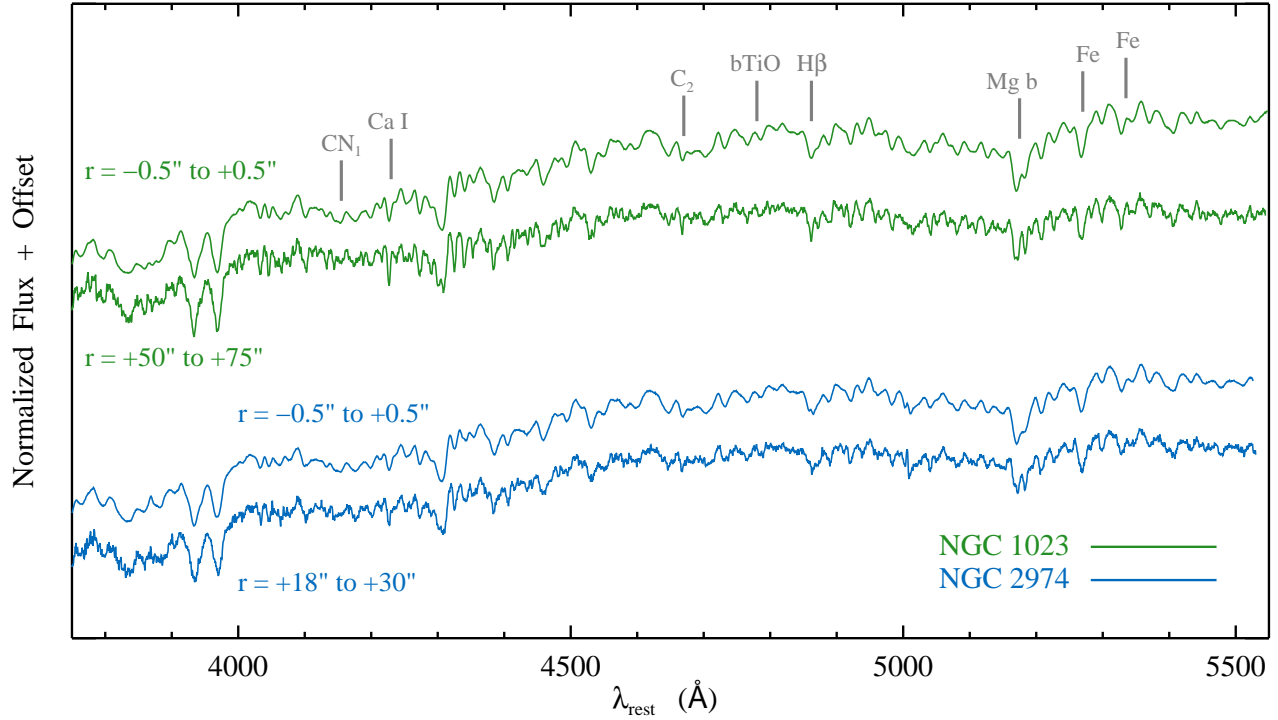,width=6.8in}
 \caption{
 LRIS blue arm spectra extracted from the central and outer regions of NGC 1023 (top, green) and NGC 2974 (bottom, blue), and shifted to rest-frame wavelengths.  Each spectrum has been sky-subtracted and flux-calibrated.  Emission lines have been removed, as described in Appendix~\ref{app:gas}.  The spectra are normalized to the same median value, and constant offsets have been added for clarity.  Stellar absorption features analyzed herein are labeled.
 }
\label{fig:bluespec}
\vspace{0.15in}
\end{figure*}

\begin{figure*}[!h]
\centering
\vspace{-0.2in}
  \epsfig{figure=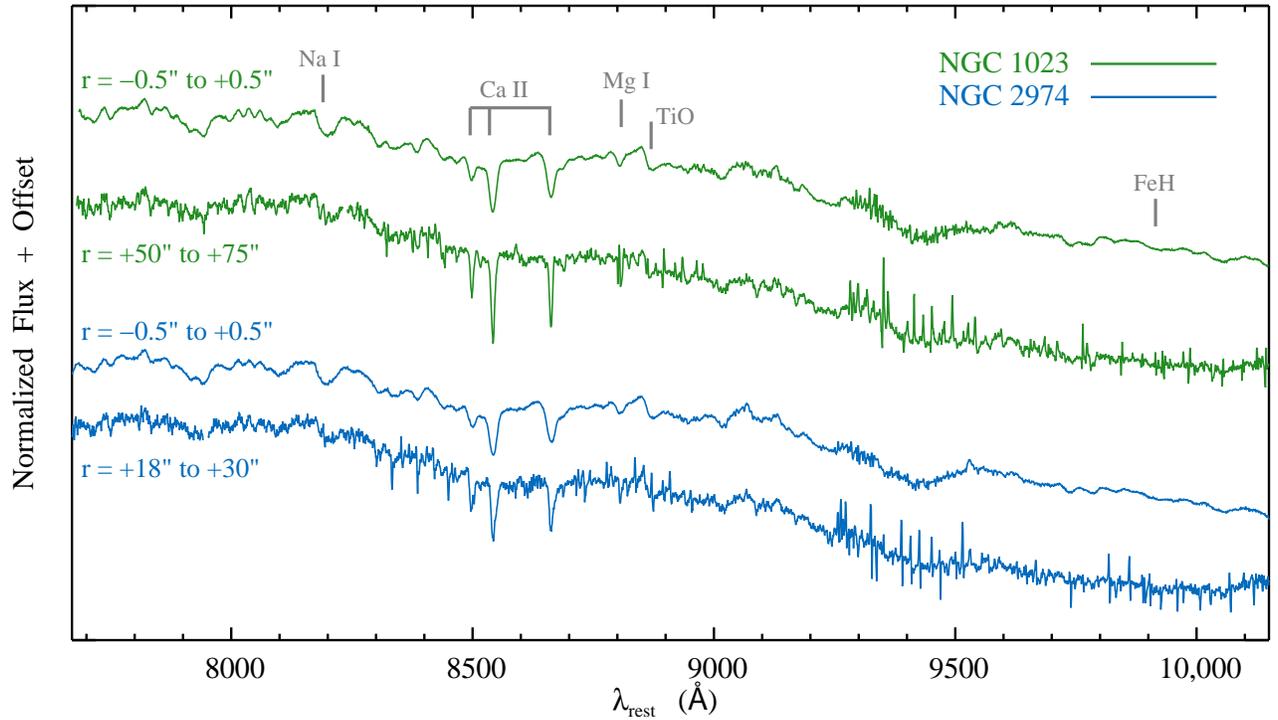,width=6.8in}
 \caption{
 LRIS red arm spectra extracted from the central and outer regions of NGC 1023 (top, green) and NGC 2974 (bottom, blue), and shifted to rest-frame wavelengths.  Each spectrum has been background-subtracted with a scaled sky spectrum, corrected for telluric absorption, and flux-calibrated.  The spectra are normalized to the same median value, and constant offsets have been added for clarity.  Stellar absorption features analyzed herein are labeled.  The increased noise at 9200-9700 \AA\, corresponds to strong telluric absorption bands.
 }
\label{fig:redspec}
\end{figure*}

Our data reduction procedures largely follow the template described in \citet{vDC12}.  The notable exception is sky subtraction.  Whereas \citet{vDC12} aligned the slit with the minor axis of each galaxy and used the far edges of the slit for sky subtraction, we wish to extract spectral features over the entire slit.  To this end we recorded separate, non-concurrent sky exposures.  On the red side, our sky subtraction requires careful wavelength calibration for each science frame, and a scaled subtraction procedure that adjusts the relative strengths of telluric emission line families, which vary on timescales of a few minutes.  Details of these various calibration steps are described below, and representative cleaned spectra are illustrated in Figures~\ref{fig:bluespec} and \ref{fig:redspec}.

%FIGURE - LRIS spectra
%
%

%FIGURE - LRIS spectra
%

\subsection{Instrumental Calibrations}
\label{sec:calib}
  
We use the {\tt lpipe} software package\footnote{The {\tt lpipe} IDL routines are available at http://www.astro.caltech.edu/$\sim$dperley/programs/lpipe.html.} (Daniel Perley, private communication) for bias subtraction, flatfield correction, and cosmic ray cleaning.  
This package first determines and subtracts bias levels from the overscan region of each frame.  
Next it performs flatfielding using halogen lamp exposures taken at the end of the observing night, with the slit and disperser in place.
The {\tt lpipe} package coadds individual flatfield exposures, computes a boxcar-smoothed response spectrum for each row of the coadded flatfield, and divides each row by its matching response spectrum to produce a pixel-to-pixel flat.
This flatfielding correction retains response variations along the spatial axis of the CCD, which are calibrated during the sky subtraction step (\S\ref{sec:sky}).
After flat-fielding we split the raw frames into the separate CCD chips for each arm and process each chip independently.  

On the red side, we use telluric OH emission lines to define a wavelength solution for each frame, 
using the IRAF routines {\tt identify}, {\tt fitcoords}, and {\tt transform}.
We divide each frame into blocks of 40 rows ($5.4''$) and extract a one-dimensional spectrum for each block.  We fit a fifth-order polynomial to the peaks of the OH lines, converting pixels to \AA.  To calibrate for wavelength variation across each chip, we then fit a two-dimensional polynomial solution to the set of peak locations from all blocks.  The 40-row extraction window permits reliable sky line fitting except for one or two blocks near the center of each galaxy; these blocks are masked from the two-dimensional fit.

The blue side includes a faint telluric line from NI at 5199 \AA\, but is otherwise devoid of telluric emission features.  We therefore derive a two-dimensional wavelength solution from daytime arc lamp frames, and assume that temporal variation in the wavelength solution can be described by a constant offset term at all wavelengths, along the entire slit.  In some frames, the NI line is too faint to perform a useful fit.   We therefore use the Fe5270 absorption feature at the galaxy center to measure the relative wavelength offsets across a sequence of frames.  We anchor the frame-to-frame offsets to a single exposure for each galaxy, where the NI 5199 feature is strong enough to establish the absolute wavelength scale.
In order to avoid substantial velocity structure when measuring the wavelength shifts in Fe5270, we only extract a $1.5''$ region at the galaxy center.  We account for each galaxy's rotation curve once we have extracted final one-dimensional spectra for all spatial bins, and before we measure line indices (see \S\ref{sec:indices}).

\subsection{Sky Subtraction and Telluric Correction}
\label{sec:sky}

The background spectrum at red wavelengths is dominated by telluric emission lines, whose relative strengths vary on timescales comparable to our exposure times.  Our goal of extracting galaxy light over the full length of the LRIS slit prohibits in-frame sky subtraction, and our sky frames must be corrected for variation in the line strengths.  We extract a high-$S/N$ sky spectrum from each science frame by applying a large aperture, offset from the galaxy center by at least $30''$.   We collapse the entire sky frame into a one-dimensional spectrum and use the {\tt Skycorr} routine \citep{skycorr}
\footnote{{\tt Skycorr} and {\tt SkyCalc} are available from the European Southern Observatory at http://www.eso.org/sci/software/pipelines/skytools}
to perform scaled sky subtraction on the spectrum extracted from our science frame.  {\tt Skycorr} adjusts the relative amplitudes of OH and O$_2$ emission groups in the input sky spectrum to best match the input science-frame spectrum, and subtracts the rescaled sky spectrum to output a clean science spectrum.

\begin{figure}[!t]
\centering
  \epsfig{figure=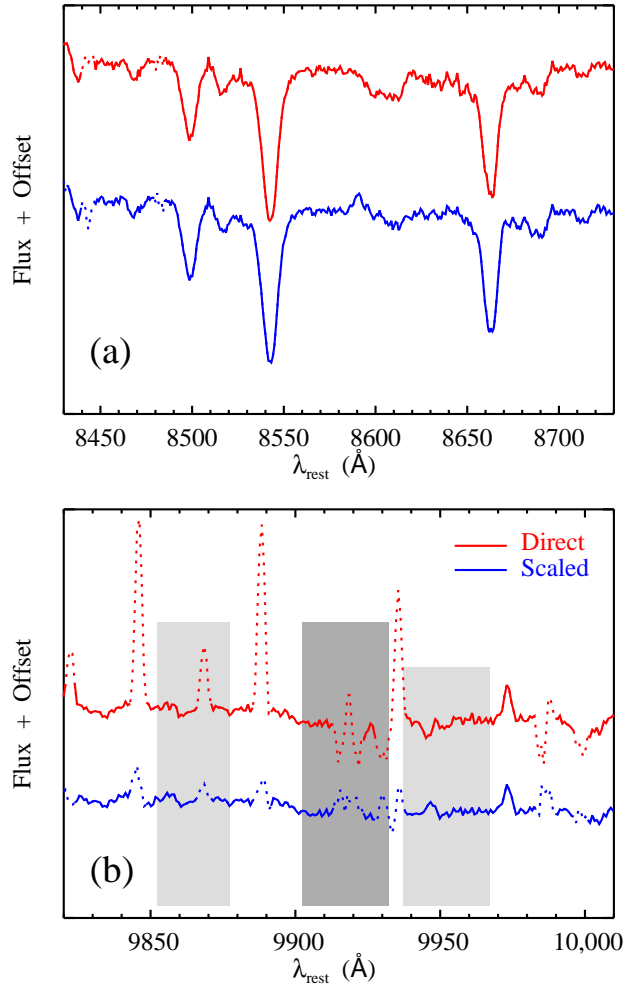,width=3.3in}
 \caption{
Sky-subtracted spectra extracted from $+30''$ to $+50''$ with respect to the center of NGC 1023.  \textit{Top:} region near the $\cat$ feature.  \textit{Bottom:} region near the $\feh$ feature, with dark and light shaded areas corresponding to the line center and pseudo-continuum regions.  In both panels, the red (upper) spectrum was obtained by directly subtracting sky frames from science frames, whereas the blue (lower) spectrum is our final spectrum after scaled sky subtraction.  Even after scaled sky subtraction, pixels corresponding to bright sky lines exhibit excess noise.  The dashed regions in each spectrum indicate areas that we masked before computing line indices.
 }
\label{fig:skysub}
\vspace{0.15in}
\end{figure}

Subtracting the {\tt Skycorr} output spectrum from the initial science-frame spectrum yields a one-dimensional ``master'' sky spectrum for that particular frame.  Next we use the halogen flats to compute the average response function of the CCD chip in the spatial dimension.  We expand the master sky spectrum as a two-dimensional array, scaled by the spatial response function, and subtract this array from the science frame.
Figure~\ref{fig:skysub} illustrates the improvement in sky subtraction after performing this scaled sky procedure, relative to direct sky subtraction.

Residuals from telluric emission lines are our dominant source of noise near the $\feh$, MgI0.88, and $\tio$0.89 features (see Table~\ref{tab:linedefs} for definitions).
These residuals are likely a combination of high shot noise from the bright lines, and wavelength calibration errors with magnitude $< 0.1$ \AA.  We have experimented with multiple variants of our sky subtraction procedure, including zoomed-in wavelength calibration over a narrow wavelength interval near the $\feh$ band, fitting the wavelength solution with different polynomial orders, and applying {\tt Skycorr} over narrower chunks of wavelength space.  None of these attempts yielded clear improvements.

%FIGURE - LRIS sky subtraction
%

In addition to their native pixels, residuals from bright sky lines may contaminate adjacent pixels during smoothing, which we employ to bring spectra to a common $\sigma$ (\S\ref{sec:indices}). 
To mitigate this, we perform a version of the iterative masking procedure described by \citet{vDC12}.  We first construct a mean sky spectrum for our observing night and flag the pixels corresponding to the brightest sky lines.  We then mask these pixels at the corresponding wavelengths in each galaxy spectrum, such that their values are interpolated from nearby good pixels before smoothing. 
As the smoothing introduces undesired correlations between good pixels and masked pixels, we return to the un-smoothed spectrum and replace only the masked pixels with the smoothed output.  We then perform the smoothing again, and iterate the substituting and smoothing steps five times.  Although pixels near the edges of our masked regions are still over-weighted in our final spectrum, the iterations serve to distribute the excess weights more broadly.  

In Figure~\ref{fig:skysub}b it is evident that the sky masking removes a substantial fraction of the $\feh$ line region at 9902-9932 \AA.  In Appendix~\ref{app:altna} we examine additional variants of the $\feh$ index that extend the line region to 9962 \AA.  These variants contain a larger fraction of unmasked pixels and yield similar radial trends in the $\feh$ index for $r < 0.5 \reff$.
As a second test, we have smoothed our spectra and measured line indices without any masking.  The resulting index values do not differ significantly from the measurements we present below.  Only the masked data exhibit a marginal upturn in MgI0.88 at large radii in NGC 1023 (Figure~\ref{fig:index1023}j).
Otherwise, the radial trends in MgI0.88 and $\feh$ are qualitatively similar for masked and unmasked measurements.

After subtracting the sky emission spectrum, we correct each red arm frame for telluric absorption.  
We start with three transmission spectra from the {\tt SkyCalc} sky model 
\citep{skycalcNoll,skycalcJones}
\footnotemark[4] %\textbf{(check that footnote is same \# as Skycorr)}: 
a baseline spectrum ($T_B$) for low airmass and low water vapor, a high-airmass spectrum ($T_A$), and a high water-vapor spectrum ($T_W$).  After convolving each model spectrum to the instrumental resolution of the LRIS red arm, we form a grid of linear combinations ($T_C$):\\
\begin{equation}
T_C = T_B + \eta(T_A-T_B) + \omega(T_W-T_B) + \zeta \;.  
\end{equation}
For each exposure, we extract a spectrum from the central 
region %$10''$ 
of the galaxy, flatten it over the range 9250-9650 \AA\ by dividing out a 
fourth-order polynomial, and determine the values of ($\eta,\omega,\zeta$) that best reproduce the observed telluric absorption over the wavelength range 9310-9370 \AA.  We then divide the frames on both CCD chips by the corresponding model $T_C$.  At small radii, noise from our transmission correction dominates the galaxy spectra over the range $\sim 9200$-9700 \AA, where there are no stellar absorption features of interest.  
More importantly, a band of telluric H$_2$O overlaps the redshifted $\na$ feature in both galaxies, with a particularly strong transmission dip at 8230 \AA\, observed wavelength.  
In Appendix~\ref{app:telluric} we assess a plausible error range for the transmission spectra used to correct the $\na$ feature, and the corresponding error $\epsilon_{\rm tel}$ in our measurements of the $\na$ line index.
We find that variations of $10\%$ 
in our adopted transmission spectrum lead to errors as high as $13\%$ and $4\%$ in the $\na$ index for NGC 1023 and NGC 2974, respectively.
These errors are incorporated along with other systematic terms in the line index measurements presented below (e.g., in Figures~\ref{fig:index1023} and \ref{fig:index2974}).
We note that our empirically derived errors are much larger than the 
estimate of 0.1-0.2\% by \citet{vDC12}.

For our wavelength coverage on the blue side, the sky background is dominated by a continuum spectrum, and the background level was very stable during our observing night in December 2013.  For the corresponding blue-arm data we directly subtract a calibrated sky frame from each calibrated science frame.
Telluric absorption is negligible on the blue side.

\subsection{Position Registration}
\label{sec:trace}

For each science exposure, we trace the position of the galaxy center by fitting a two-component Gaussian profile to the central $\approx 5''$ of each galaxy, in each of 10 wavelength blocks.  The resulting trace is interpolated to all wavelengths, and one-dimensional spectra are extracted with sub-pixel spatial precision at each wavelength, approximating the flux in each pixel to be evenly distributed in the spatial dimension.  Each science exposure also includes one chip per arm that is offset from the galaxy center.  We measure the spatial gap between the on-center and off-center chip by comparing the RA and DEC header keywords for alternating dither positions to the location of the galaxy center on the corresponding chips.  For NGC 1023 and NGC 2974 the gap spans $35''$ on the red arm and $14''$ on the blue arm.  This includes a small buffer region where the data frames are trimmed during initial calibration steps.
To calibrate for wavelength-dependence of the trace on each off-center frame, we apply the trace from an adjacent exposure where the galaxy center is positioned on the corresponding chip.

\subsection{Spectral Response Calibration}
\label{sec:fluxcal}

On the red side, we follow the flux calibration procedure of \citet{vDC12}, which uses halogen flats to measure variations over small wavelength scales and calibrates large-scale variation with a white dwarf spectrum.  We extract a one-dimensional halogen spectrum for each chip, apply 10-pixel boxcar smoothing, and divide the result into the extracted spectrum of a white dwarf calibration star.   After this initial correction step, the white dwarf spectrum should approximate a $\lambda^{-4}$ power law, with deviations on $\sim 100$ \AA\, scales arising from the intrinsic shape of the halogen spectrum.  We divide the corrected white dwarf spectrum by a $\lambda^{-4}$ function and fit the residual spectrum with a 10-point cubic spline.  We then multiply the smoothed halogen spectrum for each chip by this smoothed residual profile, flattening the intrinsic source and yielding our final response curve.

On the blue side, the steep intrinsic spectrum of the halogen light source renders even preliminary calibration ineffective, and so we divide  the observed white dwarf spectrum on each chip directly by $\lambda^{-4}$.  To remove stellar features and pixel-to-pixel noise, we mask the pixels corresponding to Balmer absorption features, interpolate from the remaining wavelengths with a cubic spline, and perform 100-pixel boxcar smoothing to obtain a final response curve.  Visual inspection of our blue-side halogen spectra reassures us that there is little response variation on scales $< 50$ \AA.

\section{Measuring Line Indices}
\label{sec:indices}

While some investigations of stellar populations employ full spectral fitting
\citep[e.g.,][]{CidFernandes05,Koleva09,Kuntschner10,CvD12b,Podorvanyuk13,CGvD14,McDermid15,Posacki15,Wilkinson15},
line indices or equivalent widths of specific absorption features are useful for qualitative interpretation and can be readily applied to a number of SPS models
\citep[e.g.,][]{Trager00a,Trager00b,DThomas05,Schiavon07,Graves08,LaBarbera13,Spiniello14}.
In Table~\ref{tab:linedefs} we list the definitions of 13 line indices discussed herein.
On the blue side we track prominent indices from the Lick/IDS system, introduced by \citet{Faber85} and updated by \citet{Worthey94} and \citet{Trager98}, as well as the bTiO index from \citet{Spiniello14}.  Our wavelength coverage with the LRIS blue arm cuts off in the middle of their aTiO feature.
At near-infrared wavelengths, we adopt line index definitions
for $\na$, $\cat$, $\feh$, $\tio$, and Mg
from \citet[][hereafter CvD12]{CvD12a}.

%TABLE - line indices
%
%{table*} sets table across whole page in emulateapj5
\begin{table*}[!t]
%\begin{small}
\begin{center}
\caption{Line Index Definitions}
\label{tab:linedefs}
%\leavevmode
\begin{tabular}[b]{llcccll}  %each c is a column, [c]entered.  For lines between columns, [c|c|c|c] etc.
\hline
%Index $\;\;\;\;\;\;\;\;\;$ & Line & Blue Pseudocontinuum & Red Pseudocontinuum & Units & Ref. \\
Index & Ref.  $\;\;\;\;$ & Line & $\;\;\;\;$ Blue Pseudo $\;\;\;\;$ & $\;\;\;$ Red Pseudo $\;\;\;$ & Units $\;$ & Dependence \\

\smallskip
 & & (\AA) & (\AA) & (\AA) & \\
%\smallskip
%(1) & (2) & (3) & (4) & (5) & (6) \\
\hline 
\\
%CaII0.39 & 3898.4 - 4002.4 & 3085.4 - 3832.7 & 4019.6 - 4051.3 & \AA & 1,2,3 \\
CN$_1$ & 1 & 4142.1 - 4177.1 & $\;$ 4080.1 - 4117.6 $\;$ & 4244.1 - 4284.1 $\;$ & mag & C$\uparrow$, N$\uparrow$, O$\downarrow$, age$\uparrow$ \\
Ca4227 $\,$ (CaI0.42) & 1 (2) & 4222.2 - 4234.8 &  4211.0 - 4219.8 & 4241.0 - 4251.0 $\;$ & \AA & Ca$\uparrow$, C$\downarrow$, age$\uparrow$ \\
C$_2$4668 $\;$ (C$_2$0.47) &  1 (2) & 4634.0 - 4720.2 & 4611.5 - 4630.2 & 4742.8 - 4756.5 $\;$ & \AA & C$\uparrow$, age$\uparrow$, O$\downarrow$ \\
bTiO & 3 & 4758.5 - 4800.0 & 4742.8 - 4756.5 & 4827.9 - 4847.9 $\;$ & mag & $\teff\downarrow$, $g\uparrow$, Mg$\uparrow$, O$\uparrow$, Ti$\uparrow$, C$\downarrow$ \\
H$\beta$ & 1 & 4847.9 - 4876.6 & 4827.9 - 4847.9 & 4876.6 - 4891.6 $\;$ & \AA & age$\downarrow$, C$\downarrow$, $\alpha\uparrow$,  Fe$\uparrow$ \\
Mg b $\;$ (MgI0.52a) & 1 (2) & 5160.1 - 5192.6 & 5142.6 - 5161.4 & 5191.4 - 5206.4 $\;$ & \AA & Mg$\uparrow$, age$\uparrow$, C$\downarrow$, $\teff\downarrow$ \\
%MgI0.52b & 2 & 5163.6 - 5218.5 & 5123.6 - 5163.6 & 5218.5 - 5258.5 & \AA \\
Fe5270 $\;$ (FeI0.52) & 1 (2) & 5245.6 - 5285.6 & 5233.2 - 5248.2 & 5285.6 - 5318.2 $\;$ & \AA & Fe$\uparrow$, age$\uparrow$ \\
Fe5335 $\;$ (FeI0.53) & 1 (2) & 5312.1 - 5352.1 & 5304.6 - 5315.9 & 5353.4 - 5363.4 $\;$ & \AA & Fe$\uparrow$, age$\uparrow$ \\
%aTiO & 3 & 5445.0 - 5600.0 & 5420.0 - 5442.0 & 5630.0 - 5655.0 & mag \\
$\na$ $\;$ (NaI0.82) & 2 &  8174.8 - 8202.7  &  8167.8 - 8174.8  &  8202.7 - 8212.7 $\;$ & \AA & Na$\uparrow$, $g\uparrow$, $\teff\downarrow$, $\alpha\downarrow$, age$\uparrow$ \\
$\na_{\rm SDSS}$ & 4 &  8177.8 - 8197.7  &  8140.8 - 8150.8  &  8230.7 - 8241.7 $\;$ & \AA & Na$\uparrow$, $g\uparrow$, $\teff\downarrow$, $\alpha\downarrow$, age$\uparrow$ \\
$\cat$=Ca1+Ca2 +Ca3 \hspace{-0.1in} & & & & & & \\
Ca1  & 2,4,5 &  8481.7 - 8510.7  &  8471.7 - 8481.7  &  8560.6 - 8574.6 $\;$ & \AA & Ca$\uparrow$, $g\downarrow$, Na$\downarrow$, $\teff\uparrow$, Mg$\downarrow$, Fe$\downarrow$ \\
Ca2  & 2,4,5 &  8519.7 - 8559.6 &  8471.7 - 8481.7  &  8560.6 - 8574.6 $\;$ & \AA & Ca$\uparrow$, $g\downarrow$, Na$\downarrow$, $\teff\uparrow$, Mg$\downarrow$, Fe$\downarrow$ \\
Ca3  & 2,4,5 &  8639.6 - 8679.6  & 8616.6 - 8639.6  &  8697.6 - 8722.6 $\;$ & \AA & Ca$\uparrow$, $g\downarrow$, Na$\downarrow$, $\teff\uparrow$, Mg$\downarrow$, Fe$\downarrow$ \\
MgI0.88 & 2 & 8799.5 - 8814.5 & 8775.0 - 8787.0 & 8845.0 - 8855.0 $\;$ & \AA & Mg$\uparrow$, age$\uparrow$, $\teff$ \\
TiO0.89 & 2 & & 8832.6 - 8852.6 & 8867.6 - 8887.6 $\;$ & ratio & $\teff\downarrow$, O$\uparrow$, Ti$\uparrow$, C$\downarrow$ \\
$\feh$ $\;$ (FeH0.99) & 2 &  9902.3 - 9932.3  &  9852.3 - 9877.3  &  9937.3 - 9967.3 $\;$ & \AA & $\teff\downarrow$, $g\uparrow$, Fe$\uparrow$ \\
\hline
\end{tabular}
\end{center}
\begin{small}
Notes: 
The index definitions above use air wavelengths, as do all figures in this paper. 
%The $\na$ index discussed throughout this paper corresponds to NaI0.82 above, and $\feh$ corresponds to FeH0.99.  
%The $\cat$ index is equal to the sum Ca1 + Ca2 + Ca3.  
The TiO0.89 index is defined as the flux ratio for the blue pseudo-continuum divided by the red pseudo-continuum.
The last column lists some of the main atomic species and stellar atmosphere properties that influence the depth of each feature.  
For integrated light, $\teff$ reflects the relative contribution of warm vs. cool stars, surface gravity $g$ reflects the relative contribution of giants vs. dwarfs, and $\alpha$ reflects the combined abundances of $\alpha$-process elements.
Arrows indicate whether the absorption index increases or decreases with respect to an increase in the respective quantity.  For instance, $g\uparrow$ means the index becomes stronger with increasing surface gravity; i.e. the feature is dwarf-sensitive.  
These dependencies have been compiled from several previous studies \citep[e.g.][CvD12]{Serven05,Schiavon07}.
Column 2 references are: 
(1) Trager et al. 1998; 
(2) CvD12  %(2) Conroy \& van Dokkum 2012a; 
(3) Spiniello et al. 2014; 
(4) La Barbera et al. 2013; 
(5) Cenarro et al. 2001.
\end{small}
\vspace{0.2in}
\end{table*}

\citet{LaBarbera13} introduced an alternative definition for the $\na$ index, whose pseudo-continuum regions traced smaller residuals between stellar population models and galaxy spectra from SDSS.  We describe this feature, $\na_{\rm SDSS}$, in Table~\ref{tab:linedefs}. 
In Appendix~\ref{app:altna} we examine the radial behavior of $\na_{\rm SDSS}$ and three new variants of the $\feh$ index, to test for biases that might arise from overlapping absorption features or contamination by sky lines.  In brief, we find that the choice of $\na$ or $\feh$ definition does not change the essential radial trends we observe (\S\ref{sec:gradients}), or our interpretation (\S\ref{sec:disc}).

Our spatial binning scheme for each galaxy is designed to maximize $S/N$ in each aperture, while retaining the ability to compare data near $0.5 \reff$ and $1.0 \reff$.  
Between $\sim 0.2 \reff$ and $\sim 1 \reff$ our bin sizes range from $5''$ to $20''$.
At smaller radii, increased surface brightness permits us to employ much finer binning, down to seeing-limited scales of $1.0''$.
In each two-dimensional science frame we determine an aperture according to the spatial registration described in \S\ref{sec:trace}, and at each wavelength we measure the mean flux over the corresponding rows.  Our spatial mean includes 3-$\sigma$ clipping at radii where the surface brightness profile is sufficiently shallow, typically $r > 5''$.  Once spectra have been extracted from individual science frames, they are coadded via direct summation.

\begin{figure*}[!t]
\centering
  \epsfig{figure=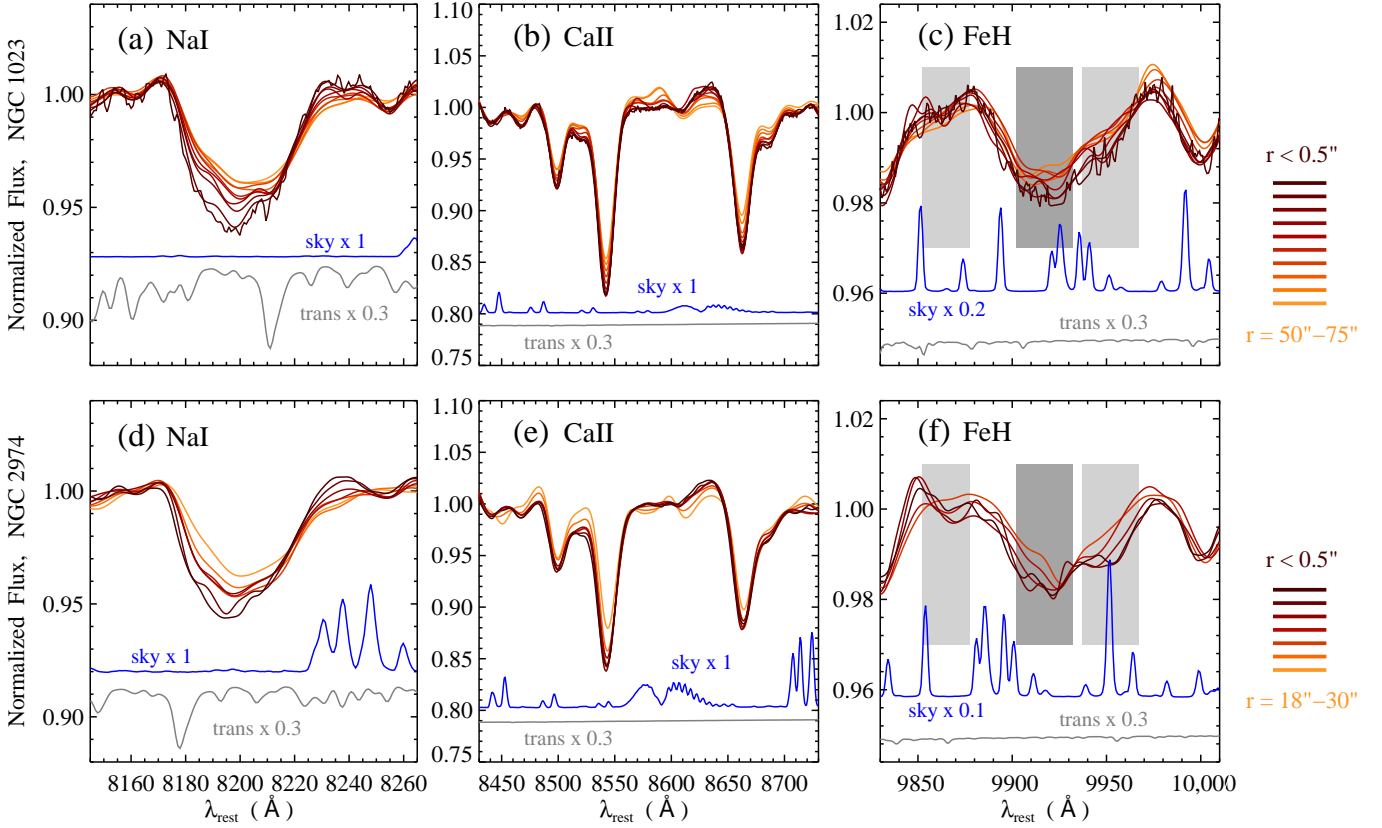,width=7.1in}
%  \vspace{0.1in}
 \caption{
Comparison of spectra at multiple radial bins in each galaxy, featuring the gravity-sensitive absorption features $\na$ (left), $\cat$ (middle), and $\feh$ (right).  For visualization purposes only, we have divided each spectrum by a third-order polynomial.  Each spectrum has been convolved to a velocity dispersion matching the central radial bin.  Top row: NGC 1023, convolved to $\sigma = 230 \kms$.  Bottom row: NGC 2974, convolved to $\sigma = 245 \kms$.  
For $\feh$, light and dark rectangles indicate the line and pseudo-continuum regions, respectively.
We also show a telluric emission spectrum (blue) and atmospheric transmission spectrum (grey), indicating the location of strong sky features.  For each galaxy, the emission spectrum is scaled to the continuum level of the central bin, with additional scalings noted in panels c and f.  The transmission spectrum is scaled by an additional factor of 0.3 in all panels.  Both sky spectra match the rest frame of the central bin.
 }
\label{fig:specrad}
\vspace{0.15in}
\end{figure*}

To assess spatial gradients in line indices, we must compare spectra with the same velocity dispersion $\sigma$, in a common rest frame.  To this end, we measure $\sigma$ and the radial velocity $v$ for each spatially binned spectrum (\S\ref{sec:gradients}), and apply a Gaussian smoothing kernel so as to artificially raise $\sigma$ to a common value.  The kernel width is chosen such that each binned spectrum has a final dispersion of $230 \kms$ for NGC 1023 and $245 \kms$ for NGC 2974, matching our highest measurement of $\sigma$ in each galaxy.  We also shift the wavelength grid for each spectrum to rest-frame wavelengths, based on our measurement of $v$.
Figure~\ref{fig:specrad} illustrates spectra near the $\na$, $\cat$, and $\feh$ features, after performing kinematic calibrations.
Our measurements of $v$ and $\sigma$ employ the {\tt pPXF} procedure by \citet{ppxf}.  In Appendix~\ref{app:sigma} we discuss possible systematic errors in measuring kinematics, and the resulting impact on our measurements of line indices.

The blue-arm spectra of NGC 2974 include strong emission lines, which must be removed before measuring the depths of nearby absorption features.  We fit and subtract an emission line component for each blue-arm spectrum by including Gaussian emission line profiles for H, NI, OII, and OIII in our list of kinematic templates for {\tt pPXF}. 
We explore uncertainties and alternative methods for the emission line fitting in Appendix~\ref{app:gas}.  
NGC 1023 shows much subtler traces of emission, which we also discus in Appendix~\ref{app:gas}. 

Once our spectra are cleaned and calibrated for kinematics, we compute equivalent widths using the formulae of \citet[][Equations 1-3]{Worthey94}, whereby the continuum level is modeled as a straight line connecting the midpoints of the red and blue pseudo-continuum bands in Table~\ref{tab:linedefs}.  Our only exception is the $\tio$0.89 index, which is expressed as the ratio of blue to red pseudo-continuum levels as defined by \citetalias{CvD12a}.

\subsection{Random and Systematic Errors}
\label{sec:errs}

In principle, noise in a galaxy spectrum can be propagated analytically to compute formal statistical errors in an ensuing equivalent width measurement.
Yet noise from sky subtraction and telluric absorption is not random and uncorrelated in our spectra.
Instead of formal error propagation, we perform a simpler and more empirical estimate of line index measurement errors, as follows.
We split the science frames for each galaxy into two subsets, with frames from each dither position evenly distributed.
We extract one-dimensional spectra from each frame, matching the spatial apertures in \S\ref{sec:gradients}, and coadd all the spectra within a subset.  After pairing data from opposite sides of the galaxy and measuring indices as described above, we have four index measurements for each interval in $r$.  We adopt the mean and standard deviation of these measurements as our final equivalent width and $1\sigma$ random error.  We then add the random errors in quadrature with the total systematic error $\epsilon_{\rm sys}$ from kinematic uncertainties, emission line removal, and telluric absorption.  We derive $\epsilon_{\rm sys}$ in Appendix~\ref{app:sys} and summarize its value for each index in Table~\ref{tab:lineerrs}.

For most of the line indices we have investigated, $\epsilon_{\rm sys}$ is comparable to the random deviations between our four subsets of data, particularly at radii $\ltsim 10''$ where our spectra have very high $S/N$.   Exceptions where random variance dominates systematic errors at all radii are the b$\tio$ index in NGC 1023, the $\cat$ index in NGC 2974, and the $\tio$0.89 and $\feh$ indices in both galaxies.  For the $\feh$ feature in particular, the noise including sky lines exceeds the $3.5\%$ total systematic error by at least a factor of two.
Our error bars in Figures~\ref{fig:index1023} and \ref{fig:index2974} below include both random and systematic terms.

The largest systematic effects occur for $\na$ and are discussed extensively in Appendix~\ref{app:sys}.  
In particular, we note that the sizable error bars for $\na$ in NGC 1023 (Figure~\ref{fig:index1023}h) are largely driven by the $\sim 10\% \; \epsilon_{\rm tel}$ term for possible errors in telluric correction.  
Biases introduced by imperfect telluric correction vary gradually with radius, as rotation in each galaxy's rotation curve shifts the $\na$ feature across the overlapping telluric band (Appendix~\ref{app:telluric}).
While this term is important for considering the absolute strength of the $\na$ index, the hypothetical telluric bias may be approximated as a uniform offset near the center of each galaxy.
Therefore, radial trends in $\na$ are more significant than suggested by Figure~\ref{fig:index1023}.  In NGC 2974, $\epsilon_{\rm tel}$ is $< 4\%$ and is outweighed by other terms in the total error budget.
When $\epsilon_{\rm tel}$ is excluded, the remaining systematic error in $\na$ ranges from 7-12\%. 

Moreover, \citet{vDC12} have suggested that the scatter in the relation between $\na$ strength and the center of the $\na$+$\tio$ blend provides a heuristic upper limit for the total error in the measured $\na$ index.  
In essence, if the only varying quantity is the $\na$ component of the $\na$+$\tio$ blend, this will drive a tight anti-correlation between the blend center and the $\na$ index strength.  Additional scatter in the relation reflects a combination of measurement errors in the $\na$ index and blend center, and independent variation of the $\tio$ component.
We have performed a linear fit to this relation for each galaxy, estimating the blend center as the wavelength of minimum flux between 8180 \AA\, and 8230 \AA.  With respect to our best fit, we find a scatter in $\na$ of 0.08 \AA\, for NGC 1023 and 0.06 \AA\, for NGC 2974.  These values are indeed comparable to our combined random and systematic errors in $\na$, which range from 0.05-0.08 \AA\, for NGC 1023 and 0.04-0.07 \AA\, for NGC 2974.  If anything, this test indicates that we have assessed our systematic errors conservatively.

%FIGURE - LRIS spectra smoothed and overlaid
%

\section{Results: Spatial Variation in Line Depths}
\label{sec:gradients}

The radial variations in the $\na$, $\cat$, and $\feh$ spectral features are illustrated in Figure~\ref{fig:specrad}, after convolving binned spectra to the same rest frame and velocity dispersion.  In both galaxies, the $\na$ and $\cat$ features become visibly shallower toward larger radius.  At $\sigma \sim 200 \kms$, the $\na$ doublet (8183 and 8195 \AA) is unresolved, and blended with a $\tio$ band at 8205 \AA.  In the left panels of Figure~\ref{fig:specrad} it is apparent that as the $\na$ blend becomes shallower toward large radii, its center shifts toward redder wavelengths, an indication that the $\na$ feature is weakening more rapidly than $\tio$. 
This trend is qualitatively consistent with variations in sodium abundance or the stellar IMF; we discuss both possibilities in \S\ref{sec:sodium}.

From Figure~\ref{fig:specrad} it is evident that the trend toward shallower $\feh$ at large $r$ is present in both the line region and the red pseudo-continuum region defined by \citetalias{CvD12a}.  While the \citetalias{CvD12a} definition highlights the deepest part of the $\feh$ bandhead, their red pseudo-continuum still includes contributions from $\feh$ and possibly $\tio$.  To ensure that radial variations in the pseudo-continuum near $\feh$ are not dominating our index measurements, we have tested three alternative variants of the $\feh$ index, including two that extend the line region to 9962 \AA.  Details are provided in Appendix~\ref{app:altna}.  All variants yield similar trends in $\feh$ with respect to $r$ near the centers of NGC 1023 and NGC 2974.

%FIGURE - Line indices vs. r for NGC 1023
%
\begin{figure*}[!p]
\centering
  \epsfig{figure=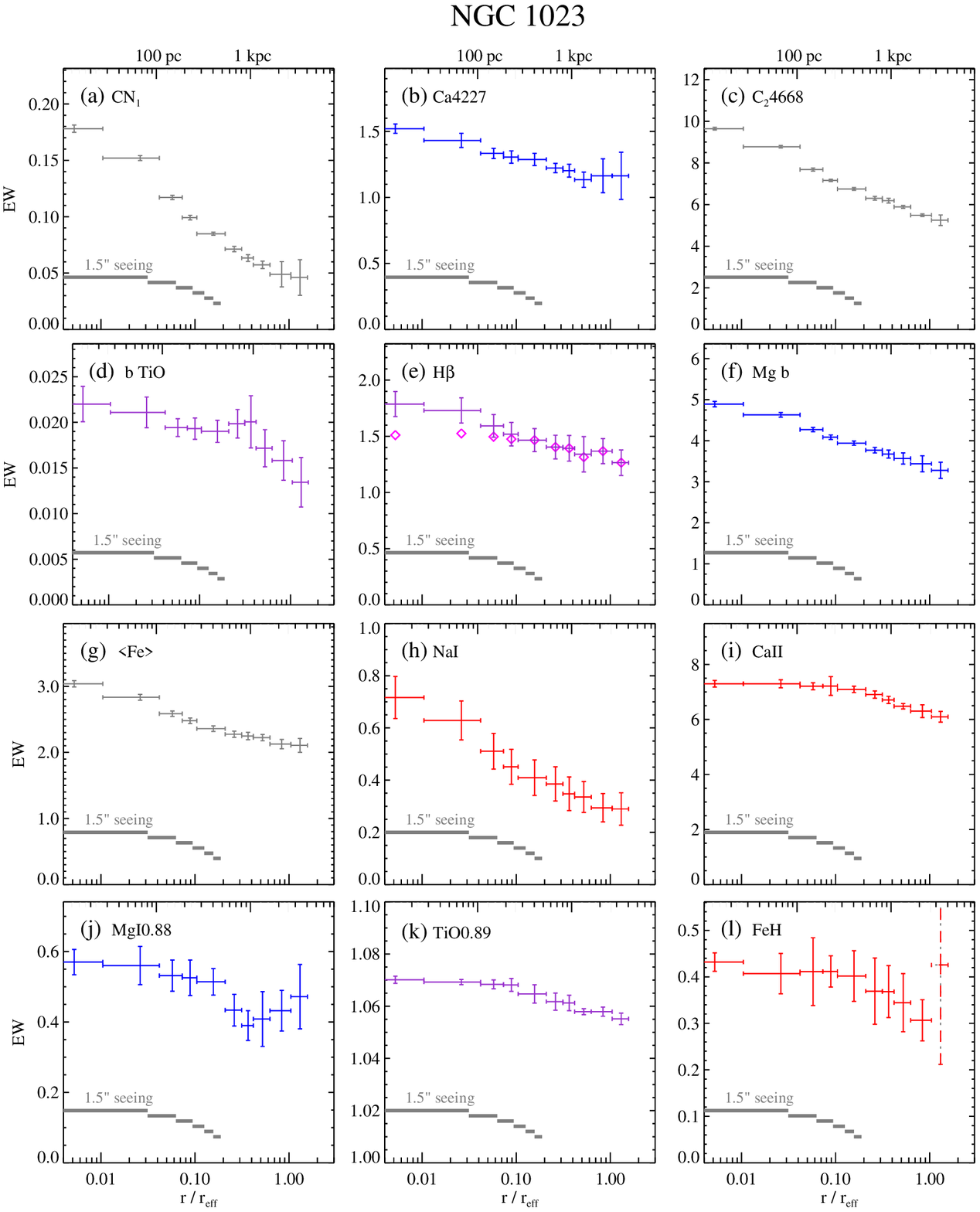,width=6.8in}
 \caption{
Selected line index strengths vs. radius in NGC 1023, corresponding to $\sigma = 230 \kms$.  Vertical errors include systematic effects discussed in Appendix~\ref{app:sys}, and horizontal error bars indicate the radial bin sizes.  Panels are ordered by wavelength and color-coded according to each index's primary use as a diagnostic:  grey (panels a, c, and g) for indicators of C, N, and Fe abundance; blue (panels b, f, and j) for $\alpha$-element indicators; purple (panels d, e, and k) for stellar age and effective temperature indicators; and red (panels h, i, and l) for IMF indicators.  However, we stress that simultaneous modeling of multiple indices is necessary to quantitatively assess the contributions of star formation history, IMF, and abundances of individual elements.  Most panels have units of \AA\, for equivalent width, except for  CN$_1$ and b$\tio$ (magnitudes), and $\tio$0.89 (ratio of blue to red pseudo-continuum).  %The $\feavg$ index is the mean of Fe5270 and Fe5335 \citep{Trager00a}.  
The $y$-axis in each panel scales from 0 to 1.3 times the maximum line depth.  
Magenta diamonds in panel e represent H$\beta$ measurements extracted without any correction for emission lines.
The last data point in panel l ($\feh$, dashed error bar) is severely compromised by sky emission.  Tabulated index values are available as online data.
 }
\label{fig:index1023}
%\vspace{0.15in}
\end{figure*}
%

%FIGURE - Line indices vs. r for NGC 2974
%
\begin{figure*}[!p]
\centering
  \epsfig{figure=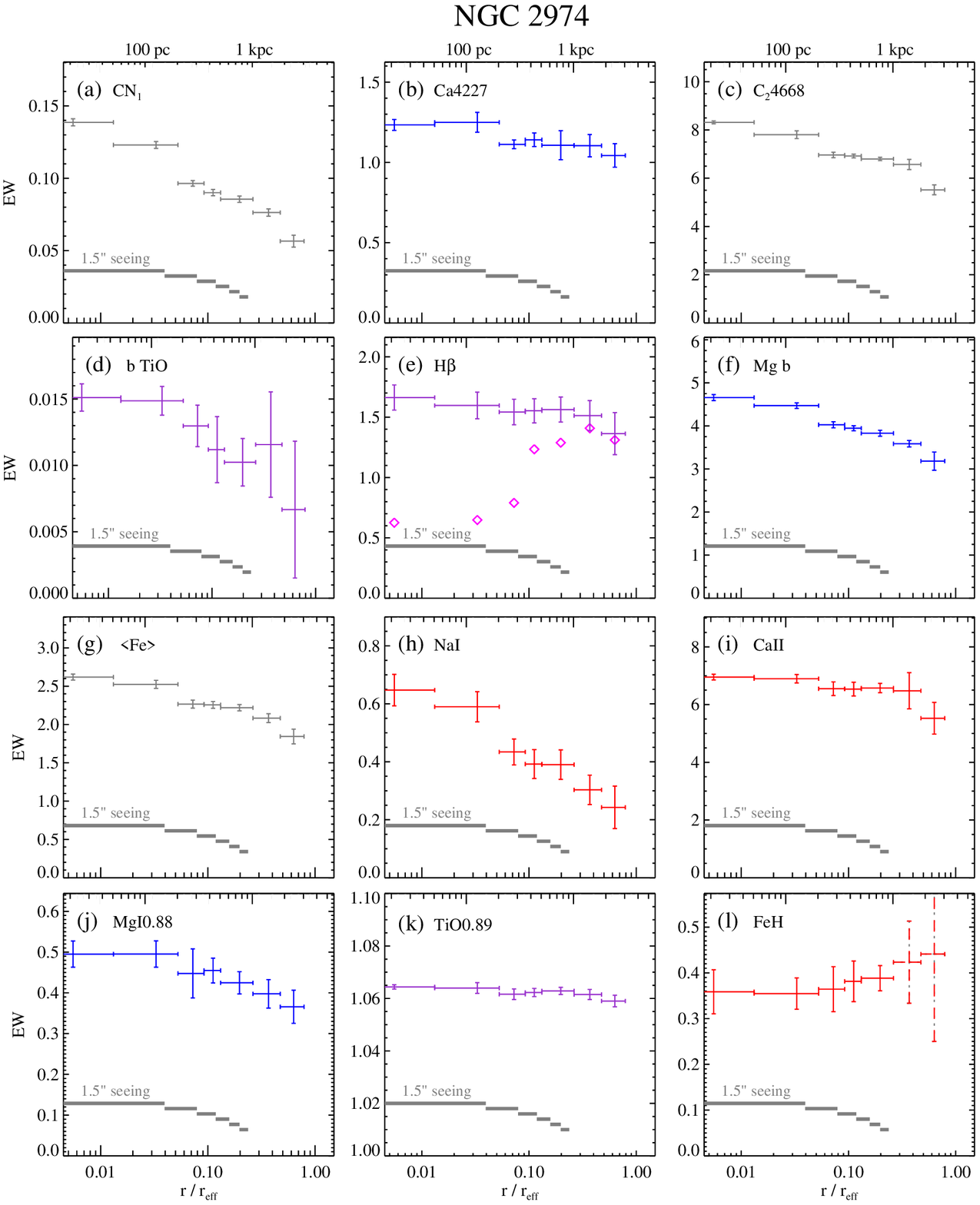,width=6.8in}
 \caption{
Selected line index strengths vs. radius in NGC 2974, corresponding to $\sigma = 245 \kms$.  Color codings and equivalent width units are the same as in Figure~\ref{fig:index1023}.  The $y$-axis in each panel scales from 0 to 1.3 times the maximum line depth.  The last two data points in panel l ($\feh$, dashed error bar) are severely compromised by sky emission.   Tabulated index values are available as online data.
 }
\label{fig:index2974}
\vspace{0.4in}
\end{figure*}

Radial trends in measured line indices are shown in Figure~\ref{fig:index1023} for NGC 1023 and Figure~\ref{fig:index2974} for NGC 2974.  
We group the H$\beta$, b$\tio$, and $\tio$0.89 indices as indicators of age and temperature; 
%(purple);
the CN$_1$, C$_2$4668, and $\feavg$ indices as indicators of C, N, or Fe abundance 
%(grey);
\footnote{We define $\feavg \equiv 0.5$(Fe5270 + Fe5335), following \citet{Trager00a}.};
the Mg~b, MgI0.88, and Ca4227 indices as indicators of $\alpha$-process elements; 
%(blue);
and the $\na$, $\cat$, and $\feh$ indices as IMF-sensitive indicators. % (red).    
Nonetheless, we stress that variations in the underlying stellar population have degenerate effects on multiple indices, and no single index or set of indices maps directly to a single stellar population property.  In particular, we will inspect the meaning of $\na$, $\cat$, and $\feh$ more carefully in \S\ref{sec:sodium}.

All of the absorption features (except $\feh$ in NGC 2974) weaken toward large radii, as expected for galaxies harboring metallicity gradients.  Yet there are noteworthy differences between the rate of decline for different indices.  
The $\na$ and CN$_1$ features exhibit steep gradients that appear nearly constant in log($r$), while other features decline less steeply on average, and/or turn over toward a flat profile in the central kpc.
Interestingly, the Wing-Ford $\feh$ band does not mirror the steep radial trend in $\na$, although both are dwarf-sensitive features.
In both galaxies, our measurements of $\feh$ are consistent with uniform strength out to $0.2 \reff$.  Beyond this radius, NGC 1023 shows a gradual decline in $\feh$ strength.  NGC 2974 shows subtle evidence for increasing $\feh$ strength toward large radii, but our outermost points for this galaxy are badly contaminated by telluric emission. 
The Ca4227, Mg~b, and $\feavg$ indices all exhibit similar radial behavior, with a decline of 15\%-25\% per dex in $r$.
Our trends in H$\beta$, Mg~b, and $\feavg$ are broadly similar to those measured by \citet{Kuntschner06} with SAURON integral-field data out to $r \approx 20''$.  
One exception is our measurement of increasing H$\beta$ toward the center of NGC 1023.  The absence of an H$\beta$ gradient in the SAURON map may reflect differences in our respective methods for removing emission lines of ionized gas.  Also, our measured Mg~b values in both NGC 1023 and NGC 2974 vary more steeply than the average trends displayed by \citet{Kuntschner06}, though noise in their two-dimensional maps hinders a direct comparison.
 
\begin{figure*}[!t]
\centering
  \epsfig{figure=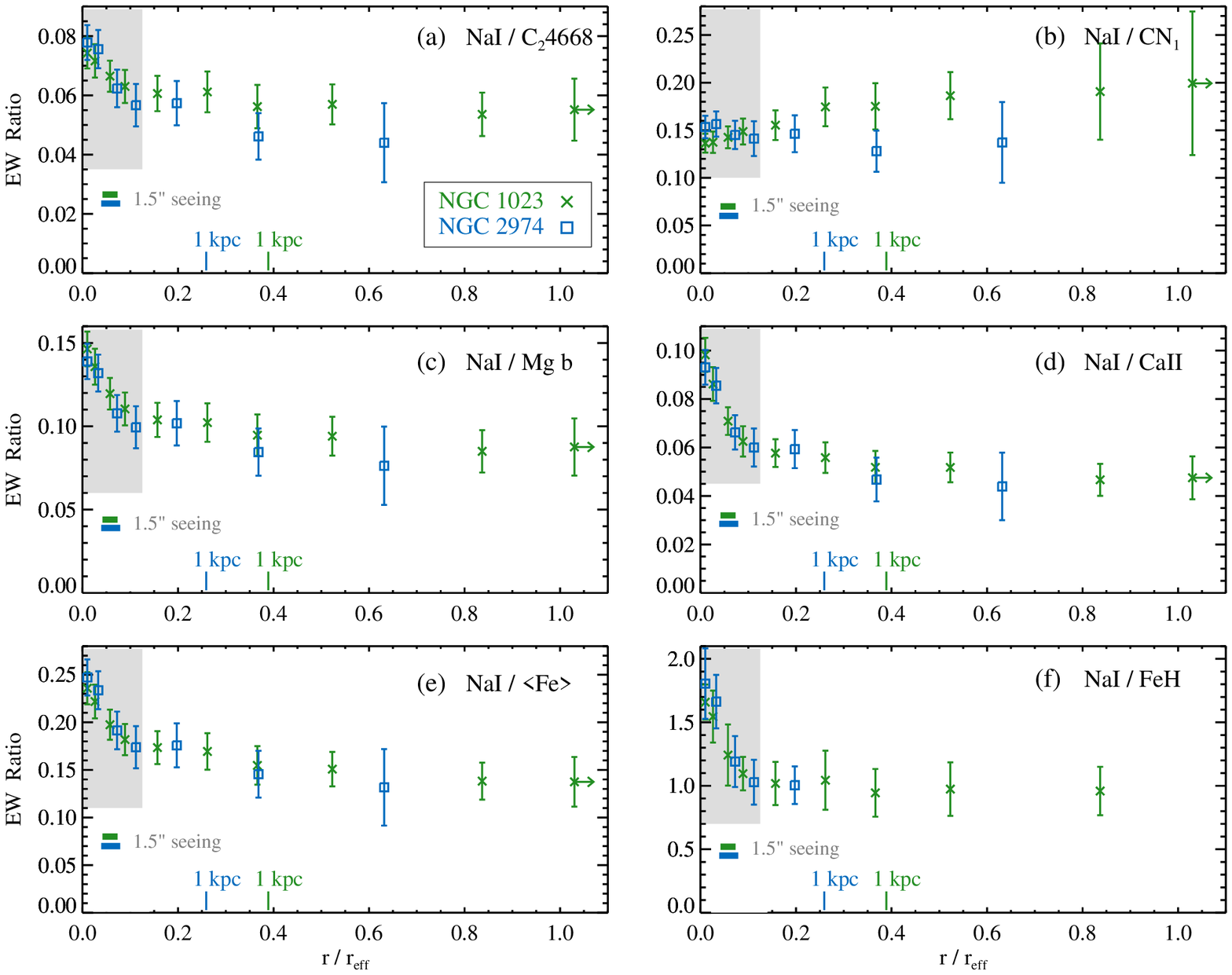,width=7.2in}
 \caption{
Ratio of $\na$ to other selected line indices as a function of radius in NGC 1023 (green) and NGC 2974 (blue).  The spatial binning is the same as Figures~\ref{fig:index1023}-\ref{fig:index2974}, but plotted on a linear scale.  In panels a-e, the outermost spatial bins extend to $1.6\reff$ for NGC 1023 and $0.8\reff$ for NGC 2974.  
For panel f ($\na/\feh$), the outermost bins extend to $1.0\reff$ for NGC 1023 and $0.26\reff$ for NGC 2974.  Grey shaded areas indicate the scale of the single-aperture measurements by \citet{vDC12}, corresponding to a physical radius of 320 pc in NGC 1023 and 480 pc in NGC 2974.  Error bars include systematics from measuring kinematics and removing emission lines, but do not include a component for telluric correction of $\na$ (see Appendix~\ref{app:telluric}).  For panel b, we measured the CN$_1$ equivalent width in \AA\, before comparing to $\na$.  
}
\label{fig:nacat}
\vspace{0.15in}
\end{figure*}

\begin{figure*}[!t]
\centering
  \epsfig{figure=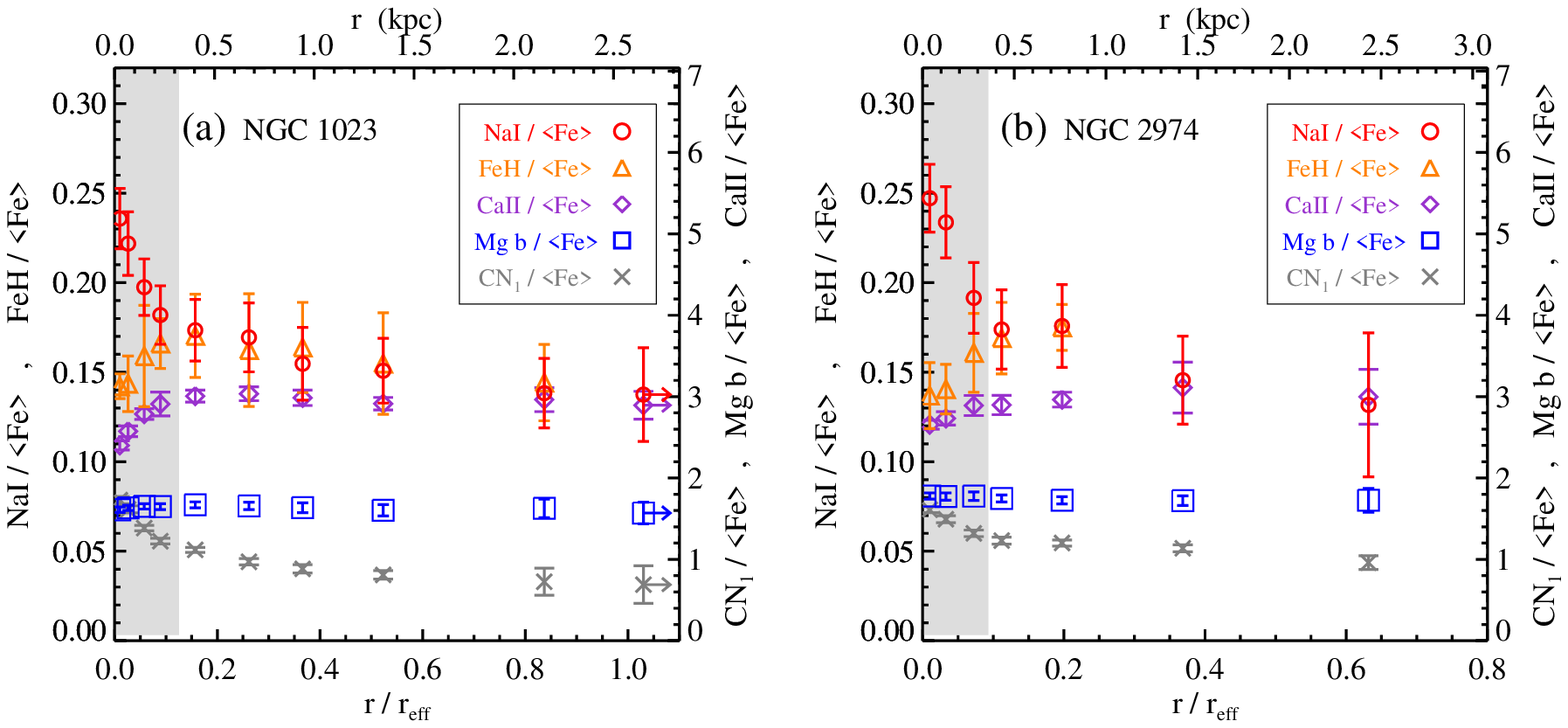,width=7.2in}
 \caption{
Ratio of IMF-sensitive indices ($\na$, $\cat$, and $\feh$) to $\feavg$ as a function of radius in (a) NGC 1023 and (b) NGC 2974.  We also illustrate the radial trends in Mg~b/$\feavg$ and CN$_1$/$\feavg$.  The spatial binning is the same as Figures~\ref{fig:index1023}-\ref{fig:index2974}, but plotted on a linear scale.  The outermost spatial bin in panel (a) extends to $1.6\reff$.  The grey shaded area indicates the scale of the single-aperture measurements by \citet{vDC12}.
}
\label{fig:N1023fe}
\vspace{0.1in}
\end{figure*}

We display ratios of selected line indices in Figures~\ref{fig:nacat} and \ref{fig:N1023fe}.  
%\ref{fig:N2974fe} 
For these figures, we have switched to a linear scale in radius so as to emphasize rapid changes within the central region of each galaxy.  We have also removed the telluric absorption term from our error bars in $\na$, as justified in \S\ref{sec:errs}.
In Figure~\ref{fig:nacat} we compare the $\na$ index to six other species.  In every case except for CN$_1$, the relative strength of $\na$ increases toward the galaxy center, with a particularly steep rise in the innermost $0.1 \reff$.  
This region turns out to be very similar to the $\reff/8$ aperture size used by \citet{vDC12}.

%FIGURE - Na vs. CaT for 2 galaxies
%
%

%FIGURE - various indices vs <Fe> for NGC 1023
%
%

%FIGURE - various indices vs <Fe> for NGC 2974
%
%\begin{figure*}[!t]
%\centering
%  \epsfig{figure=figure9.eps,width=6.8in}
% \caption{
%Same as Figure~\ref{fig:N1023fe}, for NGC 2974.  The outermost spatial bin extends to $0.8 \reff$.
%}
%\label{fig:N2974fe}
%\vspace{0.15in}
%\end{figure*}
%

In Figure~\ref{fig:N1023fe}
%s~\ref{fig:N1023fe} and \ref{fig:N2974fe} 
we compare the radial variation of IMF-sensitive and $\alpha$-element indices, relative to $\feavg$.  
Remarkably, the $\na$ and $\feh$ indices show opposite trends with respect to $\feavg$ in the central $0.1 \reff$.  
As in Figure~\ref{fig:nacat}e, $\na$/$\feavg$ rises dramatically in the central $0.1 \reff$ of each galaxy.
On the other hand, $\feh$/$\feavg$ decreases by $\sim 10$-20\%.  Even considering the large uncertainties in our measurements, the deviation between $\na$ strength and $\feh$ strength appears to be significant (Figure~\ref{fig:nacat}f).
As discussed further in \S\ref{sec:sodium}, this behavior can arise from a strong gradient in sodium abundance, whereas gradients in the low-mass IMF slope would cause $\na$ and $\feh$ to simultaneously increase or decrease.  Although IMF variations still may be possible within these galaxies, the opposing behavior of $\na$ and $\feh$ constrains the magnitude and functional form of the IMF variations relative to abundance variations.

The near constancy of Mg~b/$\feavg$ at all radii suggests that [$\alpha$/Fe] is uniform or varies mildly with radius.  The Mg~b index is modestly sensitive to carbon, and our inferred gradients in [C/Fe] (\S\ref{sec:nitrogen}) ultimately allow for a shallow decrease in [$\alpha$/Fe] toward large radii in each galaxy.
In contrast to Mg~b or Ca4227, Figure~\ref{fig:N1023fe} shows an abrupt downturn in $\cat$/$\feavg$ interior to $\sim 0.1 \reff$. 
This steep central trend could arise from a lower fraction of giant stars in the very center of the galaxy, or from an increase in sodium abundance \citepalias[e.g.,][]{CvD12a}.  Both effects predict a simultaneous increase in $\na$ strength.  
In \S\ref{sec:sodium} we argue that sodium abundance drives the radial variations in $\na$ and $\cat$.

While the $\na$ gradients in NGC 1023 and NGC 2974 are largely consistent, the CN$_1$ index shows
mild discrepancies between the two galaxies.  The overall gradient in CN$_1$ is approximately $-0.06$ mag per dex in $r$ for NGC 1023, versus $-0.04$ mag dex$^{-1}$ for NGC 2974.  This difference corresponds to a shallow decrease in $\na$/CN$_1$ toward the center of NGC 1023, versus a shallow increase for NGC 2974 (Figure~\ref{fig:nacat}b).

\section{Comparison with SPS Models}
\label{sec:CvDmodel}

To complement our qualitative interpretation of the line index gradients in NGC 1023 and NGC 2974, we compare our measurements to model spectra by \citetalias{CvD12a}.
Though rigorous fitting is reserved for future work, 
our comparison serves to highlight cases where strong radial variations in absorption line indices reveal underlying changes in abundance ratios or the stellar IMF.

\begin{figure*}[!t]
\centering
  \epsfig{figure=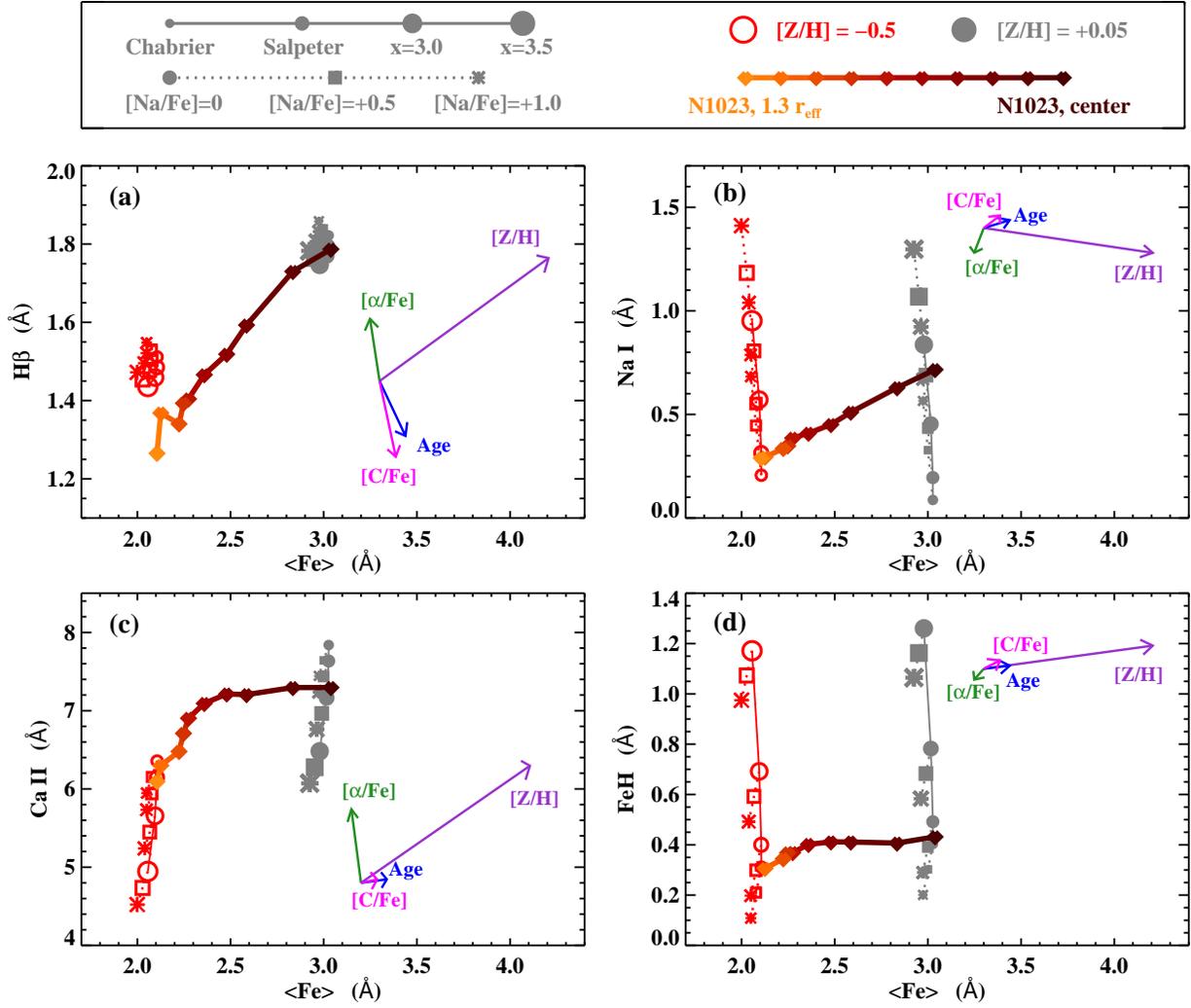,width=6.5in}
  \vspace{-0.1in}
 \caption{
Index-index trends in NGC 1023 and in SPS models, comparing different indices to $\feavg$.  The thick colored line in each panel traces the radial variations in NGC 1023, with symbols marking our measurements for each radial bin.  Overplotted are two model grids from \citetalias{CvD12a}.  In each grid we vary the IMF (symbol size; solid line) and [Na/Fe] (symbol shape; dashed line).  Closed grey symbols represent models at 10.5 Gyr, [Z/H] = +0.05, [$\alpha$/Fe] = +0.3, [Ca/Fe] = +0.2, [C/Fe] = +0.4, and [N/Fe] = +0.15.  The \citetalias{CvD12a} models with these parameters have similar Lick index values to our central bin for NGC 1023.   Open red symbols represent models with [Z/H] = -0.5, corresponding to our outermost bins for NGC 1023.  For simplicity, we have not varied age, [$\alpha$/Fe], [Ca/Fe], [C/Fe], or [N/Fe] between the grey and red model grids.  The labeled vectors represent the effects of varying age from 10.5 Gyr to 13.5 Gyr, [Z/H] by +0.55 dex, [$\alpha$/Fe] by +0.2 dex, and [C/Fe] by +0.3 dex.  Our adjustments for total metallicity [Z/H] use recently updated SPS models (Charlie Conroy, private communication).
}
\label{fig:indindfe}
\vspace{0.15in}
\end{figure*}

\begin{figure*}[!t]
\centering
  \epsfig{figure=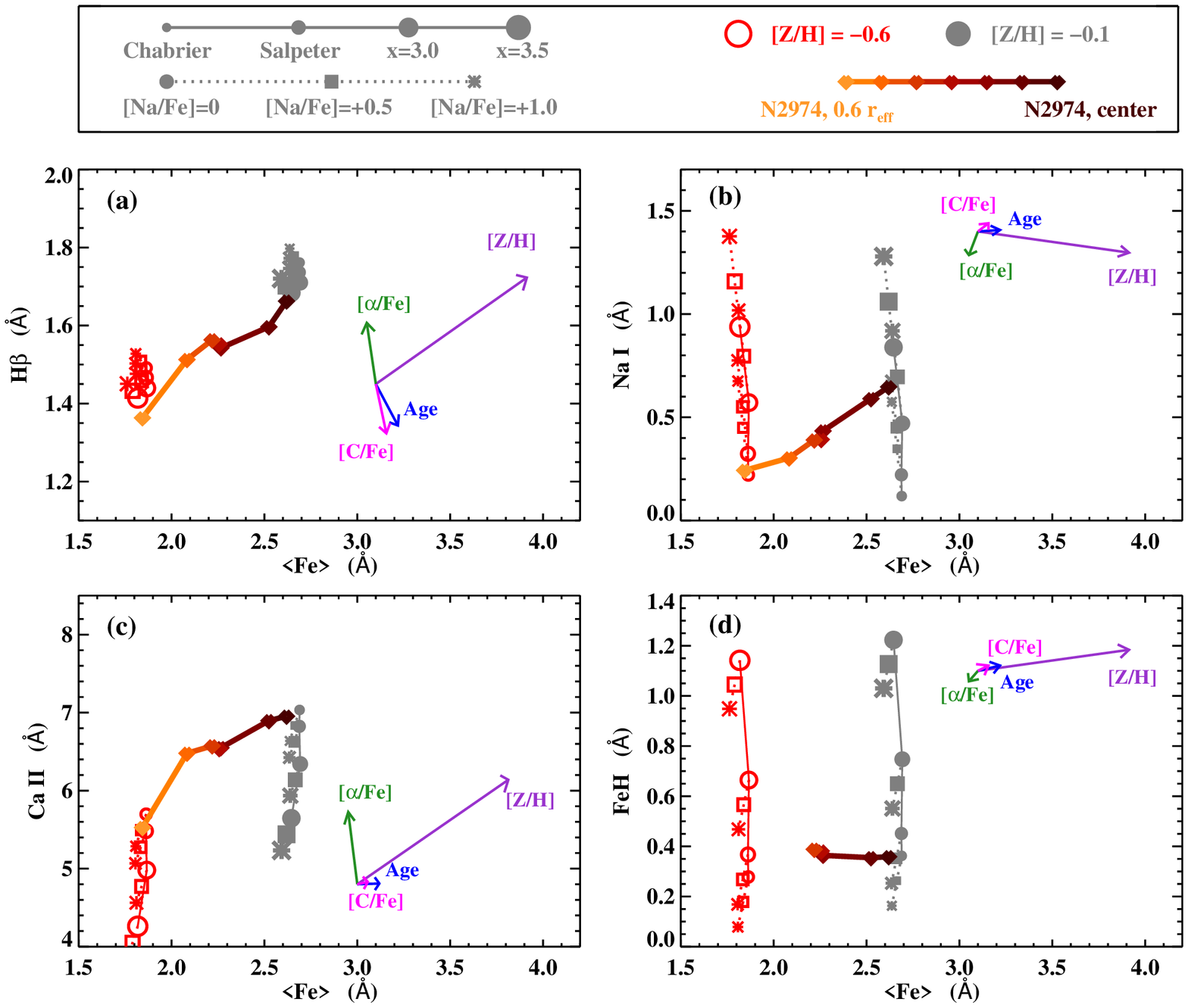,width=6.5in}
 \caption{
Index-index trends in NGC 2974 and in SPS models, comparing different indices to $\feavg$.  The thick colored line in each panel traces the radial variations in NGC 2974, with symbols marking our measurements for each radial bin.  Overplotted are two model grids from \citetalias{CvD12a}.  In each grid we vary the IMF (symbol size; solid line) and [Na/Fe] (symbol shape; dashed line).  Closed grey symbols represent models at 11.0 Gyr, [Z/H] = -0.1, [$\alpha$/Fe] = +0.3, [Ca/Fe] = +0.1, [C/Fe] = +0.3, and [N/Fe] = +0.2.  The \citetalias{CvD12a} models with these parameters have similar Lick index values to our central bin for NGC 2974.   Open red symbols represent models with [Z/H] = -0.6, corresponding to our outermost bins for NGC 2974.  For simplicity, we have not varied age, [$\alpha$/Fe], [Ca/Fe], [C/Fe], or [N/Fe] between the grey and red model grids.  The labeled vectors represent the effects of varying age from 11.0 Gyr to 13.5 Gyr, [Z/H] by +0.5 dex, [$\alpha$/Fe] by +0.2 dex, and [C/Fe] by +0.2 dex. 
}
\label{fig:indindfe29}
\vspace{0.15in}
\end{figure*}

In Figure~\ref{fig:indindfe} we illustrate the radial variation of NGC 1023 in index-index space, for H$\beta$, $\na$, $\cat$, and $\feh$ versus $\feavg$.  For comparison, each panel includes vectors indicating the isolated effects of age, total metallicity [Z/H], [$\alpha$/Fe], and [C/Fe] on model spectra.  Figure \ref{fig:indindfe29} presents the same comparisons for NGC 2974. 
In each panel of Figures~\ref{fig:indindfe} and \ref{fig:indindfe29} we display two ``families'' of models extrapolated from publicly available model spectra by \citetalias{CvD12a}.  
In our construction, a family of models samples [Na/Fe] values of 0, +0.5, and +1.0, and four IMF variants: Chabrier, Salpeter ($\alpha = 2.35$), and single (unimodal) power laws with $\alpha = 3.0$ and $\alpha = 3.5$.  
Each IMF in the \citetalias{CvD12a} models is integrated down to $0.1 \msun$.  
All parameters except for [Na/Fe] and the IMF are the same for a given family.
For each galaxy, we construct one family to approximate the line indices in our central bin, and a second family at lower metallicity to approximate the line indices at large radii.  Instead of adjusting [Fe/H] in isolation, we find better overall agreement by varying [Z/H], a parameter available in more in recent models (Charlie Conroy; private communication).

%FIGURE - Index-index plots with <Fe> as x-axis (N1023)
%
%

%FIGURE - Index-index plots with <Fe> as x-axis (N2974)
%

In order to test an appropriate range of ages, [Z/H], and other abundances, we construct a large number of model spectra based on extrapolations from the baseline grid of models by \citetalias{CvD12a}.
Starting from a given age and IMF, the abundances are applied as multiplicative response functions, originally modeled for a Chabrier IMF at 13.5 Gyr\footnote{The response function for [Z/H] in the more recent models is based on a Kroupa IMF.}.  
We extrapolate response functions from the following baseline parameters supplied by \citetalias{CvD12a}: 
%age $\in$ \{1.0, 3.0, 5.0, 7.0, 9.0, 11.0, 13.5\} Gyr;
[Z/H] $\in$ \{~-0.3, 0, +0.3~\};
[$\alpha$/Fe] $\in$ \{~0, +0.2~\};
[C/Fe] $\in$ \{~-0.15, 0, +0.15~\};
[N/Fe] $\in$ \{~-0.3, 0, +0.3~\};
[Na/Fe] $\in$ \{~-0.3, 0, +0.3~\};
and [Ca/Fe] $\in$ \{~-0.15, 0, +0.15~\}.
We then convolve each extrapolated spectrum from the native resolution of the models to $230 \kms$ or $245 \kms$, for comparison to data from NGC 1023 or NGC 2974, respectively.  We compute line indices for each model spectrum using the same procedure as for the galaxy spectra.

To estimate the stellar population parameters corresponding to a particular galaxy spectrum, we compare the indices H$\beta$, $\feavg$, Mg~b, CN$_1$, C$_2$4668, and Ca4227 to our set of extrapolated models.  These indices are relatively insensitive to variations in [Na/Fe] or the IMF.  We first find an age and total metallicity that approximately matches H$\beta$ and $\feavg$ and then iteratively adjust the age and all abundance parameters above (except [Na/Fe]) until the resulting model spectra nearly match the values of all six indices.  We emphasize that our method is not designed to produce a statistically rigorous fit, but rather to select families of models approximating the galaxy at small and large radii, so that we may visualize the effects of [Na/Fe] and the IMF on the near-infrared gravity-sensitive features.
In the following section we examine the inferred abundances and possible IMF variations.

\section{Discussion}
\label{sec:disc}

Variations in stellar masses, ages, and abundance ratios impose degenerate effects upon individual line indices.  Inferring these physical properties demands an intricate comparison between observed data and stellar population and stellar evolution models.  Herein we have attempted to present our measurements with sufficient transparency to support future analyses employing a wide range of modeling assumptions.
With the caveat that rigorous interpretation requires careful modeling, we shall discuss some qualitative trends in the relative strengths of different line indices as a function of radius, in light of previously established connections to physical stellar properties. 
We discuss gradients in age and metallicity in \S\ref{sec:age} and then examine possible origins of the steep variations in $\na$ (\S\ref{sec:sodium}) and CN$_1$ (\S\ref{sec:nitrogen}), particularly in the context of star-forming progenitors of early-type galaxies like NGC 1023 and NGC 2974 (\S\ref{sec:composite}). In \S\ref{sec:01reff} we examine each galaxy for photometric or kinematic signposts near $0.1\reff$, where the line indices and inferred stellar populations exhibit a sharp transition.  Finally, in \S\ref{sec:otherguys} we compare our findings to other recent investigations of radial IMF variations in early-type galaxies.

%FIGURE - Index-index plots with NaI as y-axis (N1023)
%
\begin{figure*}[!t]
\centering
  \epsfig{figure=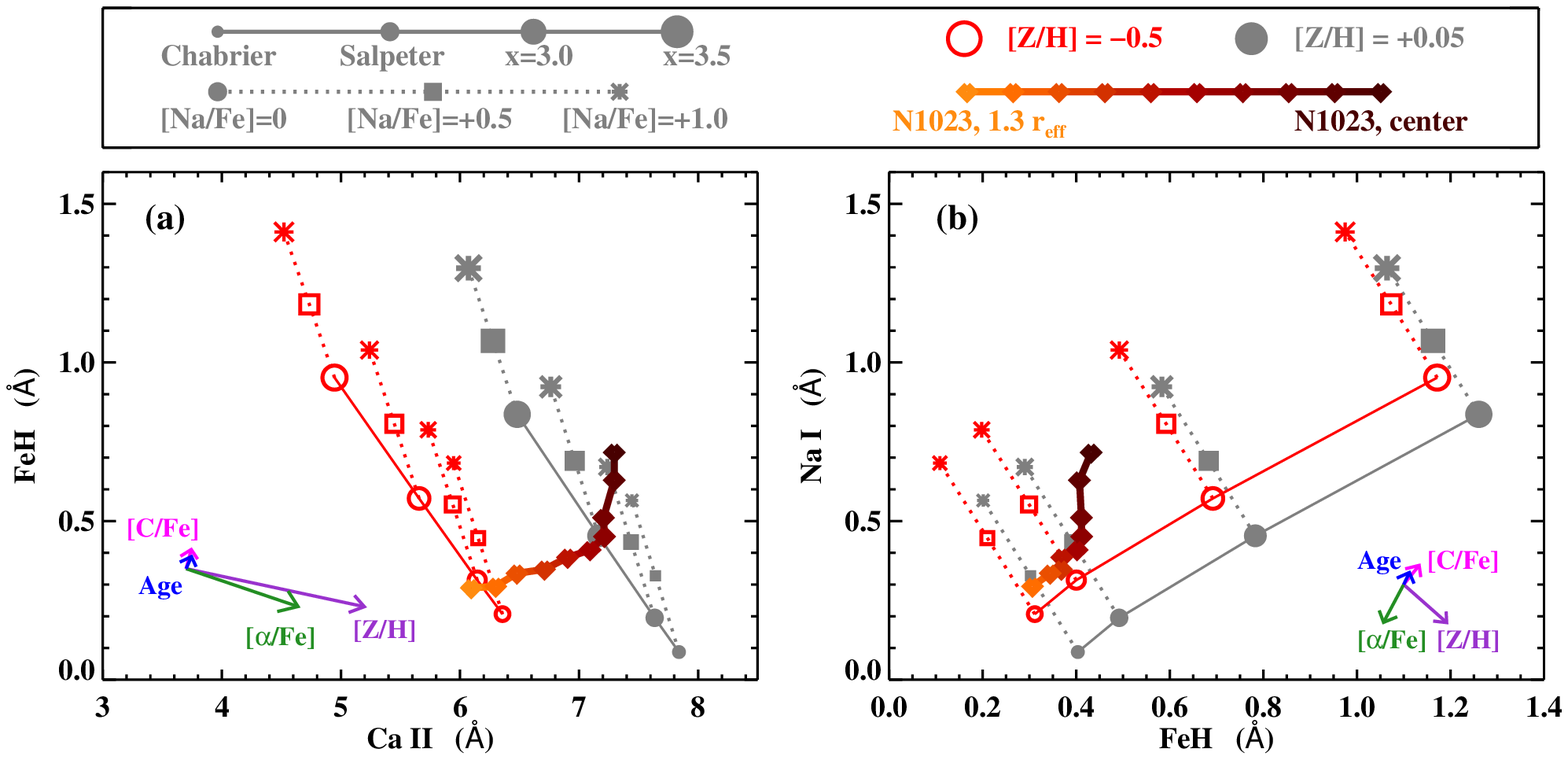,width=6.5in}
 \caption{
Index-index trends in NGC 1023 and in SPS models, comparing IMF-sensitive indices.  Data, model grids, and vectors for individual parameter variations are defined as in Figure~\ref{fig:indindfe}.  
}
\label{fig:indindna}
%\vspace{0.15in}
\end{figure*}
%

%FIGURE - Index-index plots with NaI as y-axis (N1023)
%
\begin{figure*}[!t]
\centering
  \epsfig{figure=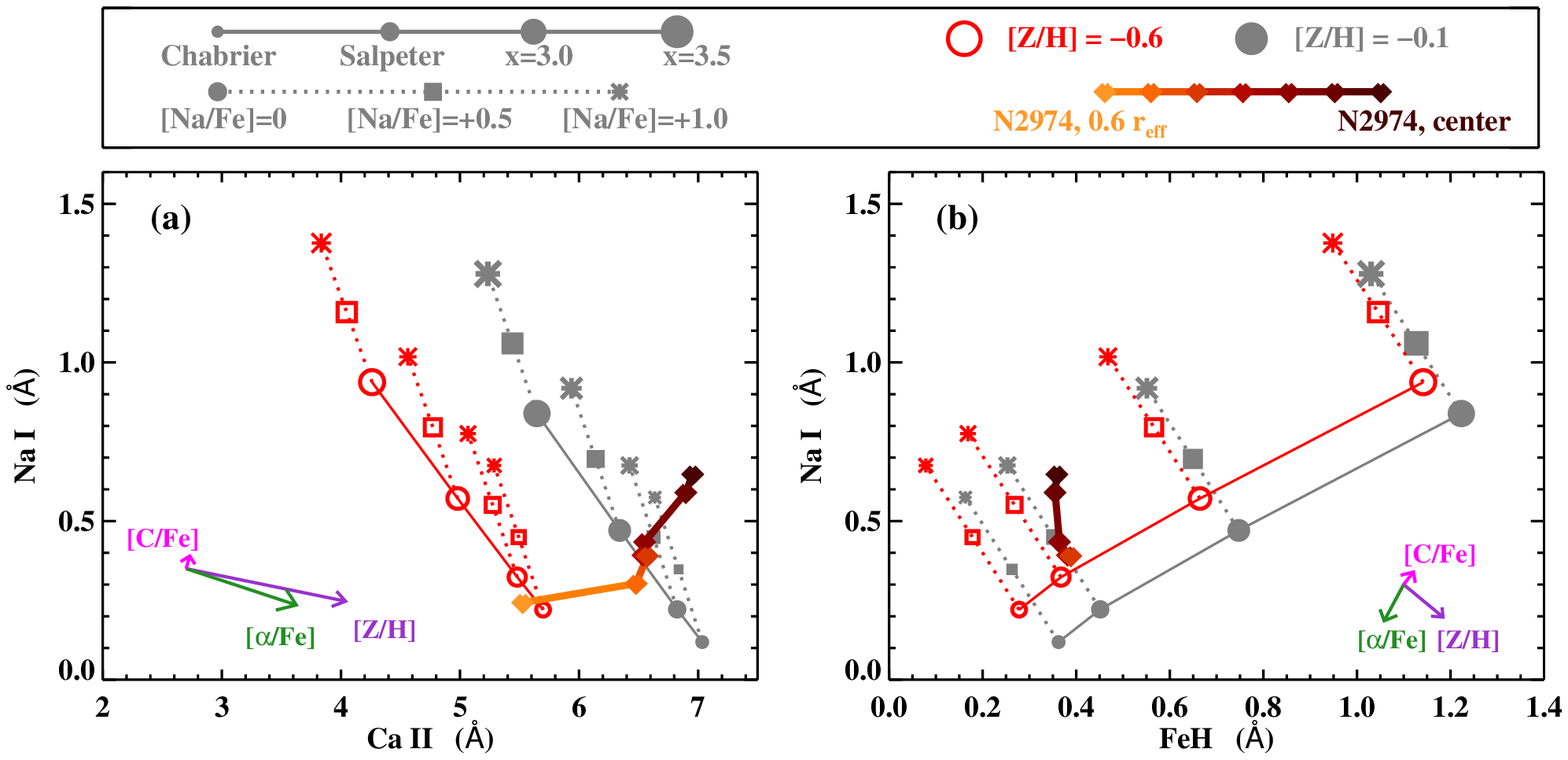,width=6.5in}
 \caption{
Index-index trends in NGC 2974 and in SPS models, comparing different IMF-sensitive indices.  Data, model grids, and vectors for individual parameter variations are defined as in Figure~\ref{fig:indindfe29}.  The age vector is not displayed, as the difference between 11.0 Gyr and 13.5 Gyr is even smaller than the [C/Fe] vector for these indices.
}
\label{fig:indindna29}
\vspace{0.15in}
\end{figure*}

\subsection{Age and Metallicity Gradients}
\label{sec:age}

Radial variations in H$\beta$ suggest age gradients in both galaxies, although the H$\beta$ index is sensitive to abundance variations as well as age (Figure~\ref{fig:indindfe}a).
Based on the process outlined in \S\ref{sec:CvDmodel}, we infer that NGC 1023 and NGC 2974 both have old stellar populations at large radii, matching 13.5 Gyr models.  
Their centers are slightly younger: $\approx 10.5$ Gyr for NGC 1023 and $\approx 11$ Gyr for NGC 2974.  As illustrated by the vectors in Figures~\ref{fig:indindfe}-\ref{fig:indindna29}, age variations in this range have a weaker impact than [Z/H] or [Na/Fe] for all indices except H$\beta$.
We note that our ages are derived after removing emission lines from the galaxy spectra.  The presence of spatially extended emission lines alongside an old stellar population is consistent with photoionization from post-AGB stars, with smaller contributions from extreme horizontal branch stars and low-mass X-ray binaries \citep{Binette94, Sarzi10}. 

Both galaxies exhibit strong gradients in total metallicity.  In NGC 1023 [Z/H] declines from $\approx +0.05$ in the central bin to $\approx$ -0.5 at $r \sim \reff$.  In NGC 2974 we measure [Z/H] $\approx$ -0.1 at the center and [Z/H] $\approx$ -0.6 at $r \sim 0.6 \reff$.  The variations in [$\alpha$/Fe] are shallower, from central values $\approx +0.3$ to outer values $\approx +0.1$ in both galaxies.  We find that [Ca/Fe] is slightly less enhanced than other $\alpha$-elements, similar to trends reported in other systems \citep[e.g.,][]{DThomas03b,Graves07,Worthey11}. %,Greene13,Greene15}.
Examination of the CN$_1$ and C$_2$4668 index strengths suggests gradients in [C/Fe] from $\approx +0.1$ at large radii to $\approx +0.4$ and $\approx +0.3$ at the centers of NGC 1023 and NGC 2974, respectively.  In contrast, we infer [N/Fe] $\approx +0.2$ at all radii.  Abundances of carbon and nitrogen are discussed further in \S\ref{sec:nitrogen}.

We caution that comparing our inferred [Z/H] values for NGC 1023 and NGC 2974 to other measurements relies on a consistent definition of total metallicity.  In the models we have employed, [Z/H] and other abundances are adjusted as independent variables, whereas in other cases [Z/H] may be a secondary quantity estimated from abundances such as [Fe/H] and [Mg/Fe].  In particular, secondary estimates of [Z/H] and [$\alpha$/Fe] depend heavily upon oxygen abundance, which is difficult to infer directly from stellar absorption features \citep[e.g.,][]{Schiavon07}.

Having approximated the ages and abundances of our inner and outer bins for NGC 1023 and NGC 2974, we can select appropriate families of models to isolate the effects of [Na/Fe] and IMF variation on various absorption indices.  
In spite of the weak trends in age, [$\alpha$/Fe], [C/Fe], and [Ca/Fe] reported above, we choose in Figures~\ref{fig:indindfe}-\ref{fig:indindna29} to illustrate grids that differ only in [Z/H].  The impacts of other abundance patterns are smaller in magnitude and are instead represented by the vectors in each figure panel.  
For both grids the adopted age, [$\alpha$/Fe], [C/Fe], [N/Fe], and [Ca/Fe] match the approximate values for the galaxy center.  With these abundances supplying a baseline, each grid then represents a family of models with varying [Na/Fe] and IMF.

We now consider two absorption indices that depart from the trend of gradual radial decline: $\na$ and CN$_1$.  As illustrated in Figure~\ref{fig:nacat}, the $\na$ index exhibits a much steeper gradient than other features in the central few hundred pc of each galaxy.  
Figures~\ref{fig:indindna} and \ref{fig:indindna29} also indicate a clear break in the behavior of $\na$ for the central 3-4 bins in each galaxy, relative to the trends in $\cat$ and $\feh$.
As discussed further in \S\ref{sec:sodium}, this break appears consistent with an abrupt rise in sodium abundance interior to $\sim 0.1 \reff$, and limits radial IMF variations to stellar masses above $\sim 0.4 \msun$.
Surprisingly, CN$_1$ is the only feature showing similar radial variation to $\na$ (Figure~\ref{fig:nacat}b).  We have considered both carbon and nitrogen abundance variations as a possible explanation for the steep gradient in CN$_1$ and discuss these possibilities in \S\ref{sec:nitrogen}.

\subsection{IMF vs. Sodium Abundance Variations}
\label{sec:sodium}

The $\na$ feature is primarily sensitive to surface temperature, surface gravity, and sodium abundance.  
The second effect makes it a strongly dwarf-sensitive feature in uniformly old stellar populations.  Thus it is tempting to interpret our observations as a steep gradient from a bottom-heavy IMF at the galaxy center to a shallower IMF slope at $r \gtsim 0.1 \reff$ ($\sim 300$ pc).  
However, this interpretation must be reconciled with the relatively mild decline in $\feh$ and the opposing behavior of $\na$/$\feavg$ and $\feh$/$\feavg$.
While the $\na$ and $\feh$ features both peak in sensitivity for stellar masses $m < 0.2 \msun$, the sensitivity of $\feh$ declines much faster toward higher masses, such that $\na$ is more than twice as sensitive to the number of stars with $m \geq 0.4 \msun$ \citepalias[see Figure 17 of][]{CvD12a}.
Thus a strong radial trend in $\na$ but not $\feh$ could reflect an IMF gradient whose ``bottom-heavy'' nature is only expressed above $\sim 0.4 \msun$.  
A bimodal IMF whose slope only varies above $\sim 0.5 \msun$ has been shown to agree with the dynamical masses of early-type galaxies \citep[e.g.,][]{LaBarbera13,LaBarbera15b,Spiniello14}.  Yet the magnitude of the variation in $\na$ would demand drastic IMF variation on sub-kpc scales if the IMF slope above $\sim 0.4 \msun$ were the sole driver.  
The b$\tio$ index also varies less steeply than $\na$, although its strong temperature-sensitivity and the overlapping Mg4780 line leave some doubt over whether it is a valid IMF indicator \citep[e.g.,][]{Serven05,LaBarbera13,Spiniello14}. 

Varying sodium abundance offers an alternative explanation for the steep variation of the $\na$ feature inside $\sim 0.1 \reff$.  
Indeed, the outer regions of both galaxies align with nearly solar [Na/Fe] in Figures~\ref{fig:indindna} and \ref{fig:indindna29}, while the central index strengths indicate [Na/Fe] between +0.5 and +1.0.
Due to its prominent role as an electron donor, sodium impacts other features via the atmospheric electron pressure.  In particular, an increase in [Na/Fe] will drive a mild decrease in ionized calcium \citepalias{CvD12a}.  This is qualitatively consistent with the flattening we observe in the $\cat$ index near the center of NGC 1023 and NGC 2974, and the corresponding decline in $\cat$/$\feavg$.  

Comparing $\na$ and $\feh$ in Figures~\ref{fig:indindna}b and \ref{fig:indindna29}b, we find that both galaxy centers occupy the same locus on the central model grid: [Na/Fe] is just below +1.0 and the unimodal IMF slope lies between 2.35 and 3.0.  The trends in $\na$ versus $\cat$ are more difficult to interpret, as IMF and [Na/Fe] variations are more degenerate for this pair of indices.   Furthermore, the $\cat$ predictions from the SPS models are sensitive to our estimate of [Ca/Fe].  Allowing for an error $\sim 0.5$ \AA\, in the horizontal placement of the model grids in Figures~\ref{fig:indindna}a and \ref{fig:indindna29}a, we again find that both galaxy centers are consistent with a sodium enhancement of approximately 1.0 dex and an IMF slope between 2.35 and 3.0.  At large radii, the $\na$ and $\cat$ strengths of both galaxies (as well as $\feh$ in NGC 1023) are roughly consistent with a Chabrier IMF.
We note that while the Chabrier form is directly employed in the models by \citetalias{CvD12a}, it is not a unimodal power law, and the apparent transition from $\alpha > 2.3$ to Chabrier may not fully reflect the unimodal or bimodal form of the IMF at different radii.

Spatially resolved measurements of the Na D feature near 5890 \AA\, would provide additional evidence for or against gradients in [Na/Fe].  Unfortunately, our settings for LRIS did not cover Na D in NGC 1023 or NGC 2974.
\citet{Jeong13} constructed a stacked SDSS spectrum of early-type galaxies with strong Na D absorption and reported that the Na D line strength requires super-solar [Na/Fe] as well as [Z/H], with IMF slope having a relatively minor effect on the Na D feature.  
Although their analysis provides circumstantial evidence in favor of [Na/Fe] driving trends in $\na$, 
their measurements leave room for IMF variations in a subset of their galaxies, which are stacked over $\sigma \sim 150$-$300 \kms$.
%their single stack including galaxies with $\sigma \sim 150$-$300 \kms$ may wash over competing trends across the galaxy population.

Sodium is produced primarily through the Ne-Na chain,
which can be activated during core burning in massive stars 
\citep[e.g.,][]{WW95,Decressin07,Kobayashi06,Kobayashi11}
or at the base of the convective envelope in intermediate-mass stars on the asymptotic giant branch (AGB), a process known as hot bottom burning
\citep[e.g,][]{CDC81,DD90,Ventura08a,Ventura08b,Karakas10}.
Although the dredge-up of heavy elements to the outer envelope of AGB stars is most efficient at low metallicities, solar-metallicity AGB stars may still produce and eject non-negligible quantities of sodium
\citep[e.g.,][]{Mowlavi99,Karakas02,Herwig05}.

In addition to IMF and [Na/Fe] variations, \citetalias{CvD12a} determined that $\na$ could be strengthened by decreasing the number of horizontal branch and AGB stars, or the number of extremely cool giants (M7III).  Yet both of these effects substantially weaken the $\tio$0.89 feature, which we observe to be nearly constant over the radii where $\na$ varies most steeply.
While our interpretation of $\na$ and other features is self-consistent within the framework of our adopted SPS models, some recent studies have cautioned against using sodium indices to investigate the IMF.
\citet{Spiniello15a} noted inconsistencies between predicted sodium depths from different SPS models, while \citet{Smith15b} found disagreement between the total mass-to-light ratios in two lensing galaxies and the IMF and [Na/Fe] values inferred from the Na D and 1.14 $\mu$m Na features.

\subsection{Carbon vs. Nitrogen Abundance Variations}
\label{sec:nitrogen}

The CN$_1$ index is sensitive to carbon and nitrogen abundance and exhibits a strong radial trend in both galaxies.  Our other carbon-sensitive feature, C$_2$4668, does not decline as steeply as CN$_1$ from $r = 0$ to $r \sim 0.1 \reff$, and thus it is tempting to interpret the trend in CN$_1$ as a decrease in nitrogen abundance.  Yet our comparison with the SPS models by \citetalias{CvD12a} suggests that a gradient in [C/Fe] is better able to reproduce the observed trends in CN$_1$ and C$_2$4668.  This counterintuitive result occurs in part because CN$_1$ is more sensitive to [Z/H], such that the metallicity gradient in each galaxy drives a seemingly exaggerated gradient in CN$_1$ strength.
Nonetheless, the absolute values of CN$_1$ relative to $\feavg$ indicate that nitrogen is mildly enhanced at all radii, with [N/Fe] $\approx +0.2$.
Neither CN$_1$ nor C$_2$4668 are sensitive to sodium abundance or IMF variations.
% \citepalias[see Figure 13 of][]{CvD12a}.  

Carbon is produced by helium fusion in the cores of stars and released into the interstellar medium (ISM) via stellar winds or Type II supernovae.  Models of massive stars ($m > 10 \msun$) produce sizable carbon yields at solar and slightly lower metallicities \citep[e.g.,][]{Meynet02b,Dray03a,Dray03b}.  In intermediate-mass stars the final carbon yield is sensitive to initial mass and metallicity, as hot bottom burning efficiently converts carbon to nitrogen.  At solar metallicity, stars with $m \approx 2.5$-$4 \msun$ transport carbon-rich material to the surface during third dredge-up events, without reaching sufficient temperatures for hot bottom burning \citep[e.g.,][]{RV81,Gavilan05,Herwig05,Karakas14,Ventura15}.
Abundance patterns in the Milky Way disk and nearby star-forming galaxies -- particularly an upturn in [C/O] toward solar metallicity -- have prompted differing interpretations of the dominant source of carbon enrichment in these environments.  
On one hand, the absolute quantities of carbon and oxygen are consistent with open-box models of chemical evolution in star-forming galaxies, employing massive star yields \citep[e.g.,][]{Gustafsson99,Henry00}.  
Conversely, observations of nearly constant [C/Fe] over $\sim 1$ dex in [Fe/H] suggest that the ISM becomes carbon-enriched on timescales similar to Type Ia supernovae, pointing to intermediate-mass stars as the main supplier \citep[e.g.,][]{Chiappini03a,Bensby06}.

Nitrogen is produced via the CN cycle.  
Whereas primary nitrogen production is driven by 
the mixture of hydrogen- and helium-burning regions in the first generation of stars,  
the nitrogen abundance of moderate- and high-metallicity stellar populations is dominated by secondary production from later generations whose interiors are already seeded with carbon.
Similar to sodium, nitrogen production may occur in the cores of massive stars or in the stellar envelopes of AGB stars above $\sim 4 \msun$
\citep[e.g.,][]{RV81,WW95,Gavilan05,Ventura08a,Ventura08b,Karakas14}.
Observations of [N/Fe] in the Galactic disk and halo 
-- including globular clusters ranging from [Fe/H] $\sim -2.5$ to [Fe/H] $\sim -0.5$ -- 
are consistent with contributions from both massive and intermediate-mass stars
(e.g., Chiappini et al. 2005; Hirschi 2007; cf. Cohen et al. 2005).

Models of fast-rotating massive stars can boost CNO yields by mixing the hydrogen shell and helium-burning core \citep[e.g.,][]{Meynet02a,Meynet02b}.
In contrast, hot bottom burning in AGB stars will drive a strong anti-correlation between [C/Fe] and [N/Fe] unless a second mechanism is responsible for producing one of these elements.  Although this anti-correlation has been observed in some globular clusters \citep[e.g.,][]{Cohen05,Ventura08b}, the radial variation of CN$_1$ and C$_2$4668 in NGC 1023 and NGC 2974 is inconsistent with opposing gradients in [C/Fe] and [N/Fe].  Rather, the trend of rising [C/Fe] at constant [N/Fe] suggests separate sources of carbon and nitrogen production.  \citet{Henry00} reached a similar conclusion after compiling data from individual stars in the Galactic disk and halo, plus Galactic and extragalactic HII regions.

\subsection{Abundance Ratios in Composite Stellar Populations}
\label{sec:composite}

The stellar populations of early-type galaxies are typically old, with high metallicities resulting from multiple generations of star formation and ISM enrichment in a deep potential well.  Their high [$\alpha$/Fe] ratios could arise from short star formation timescales or a top-heavy IMF, both of which rapidly seed the ISM with $\alpha$-process elements via Type II supernovae \citep[e.g.,][]{Worthey92,DThomas99,DThomas05}.
Here we briefly explore scenarios that could yield an excess of sodium and carbon in the centers of early type galaxies, with steep abundance gradients to larger radii.

The most straightforward hypothesis for excess sodium and carbon is a larger fraction of the stars responsible for producing these elements.  This could be the direct result of an IMF with a shallower slope above $\sim 10 \msun$ if sodium and carbon are produced in massive stellar cores.  
The form of the IMF at high densities is an ongoing challenge for models of star formation, and may be especially sensitive to the initial gas density structure and the role of turbulence in driving fragmentation
\citep[e.g.,][]{Chabrier14,Krumholz14}.  However, IMF variations cannot be invoked to explain correlated gradients in [C/Fe] and [Na/Fe] if AGB stars are the dominant source of both sodium and carbon.  This is because the IMF above $\sim 2 \msun$ sets the number of stars that experience hot bottom burning, which exerts opposite influences on carbon and sodium yields.

Another factor in the observed excesses is the amount of enriched stellar ejecta accreted onto existing and newly forming stars, during the peak of star formation.  For instance, the high stellar densities in globular clusters allow for prolific accretion onto 
existing stars, as long as the cluster is massive enough to retain low-velocity ejecta
(e.g., Renzini 2008; Conroy 2012; cf. Fenner et al. 2004). 
Using the deprojection procedure of \citet{Geb96}, we find that 
the luminosity densities of NGC 1023 and NGC 2974 rise from 2-6 $\lsun$ pc$^{-3}$ at $0.1 \reff$ to 200-400 $\lsun$ pc$^{-3}$ at $0.01 \reff$, comparable to the average densities of some globular clusters.  
An extremely high value of [Na/Fe] $\approx 1$ dex has been measured in the central 15 pc of M31 \citep{CvD12b,Zieleniewski15}, while red giants in the Galactic bulge and Galactic globular clusters both exhibit [Na/Fe] between 0 dex and +0.5 dex \citep[e.g.,][]{Lecureur07,Roediger14,Johnson14,Johnson15}.
Although support for this mechanism in globular clusters is still contentious, the rapid increase in sodium abundance toward these galaxies' densest regions is at least a plausible consequence of pollution by stellar ejecta.

Compiled measurements of sodium, carbon, and nitrogen in other early-type galaxies do little to clarify the observations reported herein.
Spectra of the central $\approx 0.6$-1 kpc in early-type galaxies exhibit stronger CN$_1$ features and steeper variation of Na D versus $\feavg$ than composite stellar population models for elliptical galaxies or the Galactic bulge \citep{Trager08,Serven10,Tang14}.
\citet{Graves07} and \citet{Johansson12} report slightly higher carbon enhancement than nitrogen enhancement in stacked spectra of SDSS galaxies, while \citet{Greene13,Greene15} report higher [N/Fe] than [C/Fe] for a combined sample of 82 galaxies.
Although stacked early-type galaxy spectra allegedly range from 0 dex to +1 dex in [N/Fe], the variations between different studies are larger than the inferred trends within any individual sample,
suggesting systematic errors in at least some measurements of [N/Fe].

\subsection{Is $0.1 \reff$ a Special Radius?}
\label{sec:01reff}

%FIGURE - NaI vs. v and sigma  (N1023 near center)
%
\begin{figure}[!t]
\centering
%\vspace{0.1in}
  \epsfig{figure=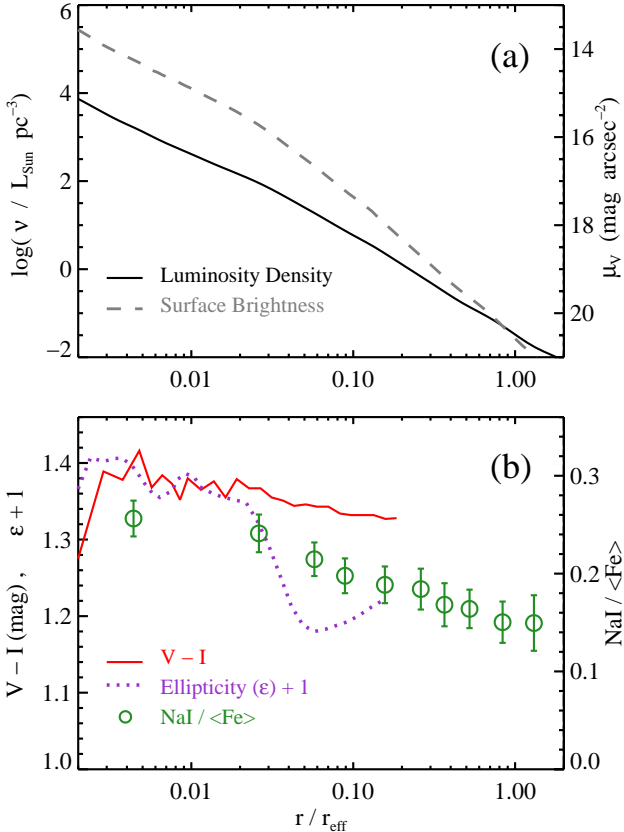,width=3.3in}
 \caption{
Major-axis photometry of NGC 1023.  (a) Surface brightness profile (dashed grey line), combining HST-based measurements by \citet{Lauer05} and ground-based measurements by \citet{Scott09}.  The solid black line illustrates the corresponding deprojected luminosity density.  (b) $V$-$I$ color and ellipticity $\epsilon$ from \citet{Lauer05}, overlaid with our measurements of $\na$/$\feavg$.
}
\label{fig:phot1023}
\vspace{0.1in}
\end{figure}

%FIGURE - NaI vs. v and sigma  (N1023 large radius)
%
\begin{figure}[!t]
\centering
%\vspace{0.1in}
  \epsfig{figure=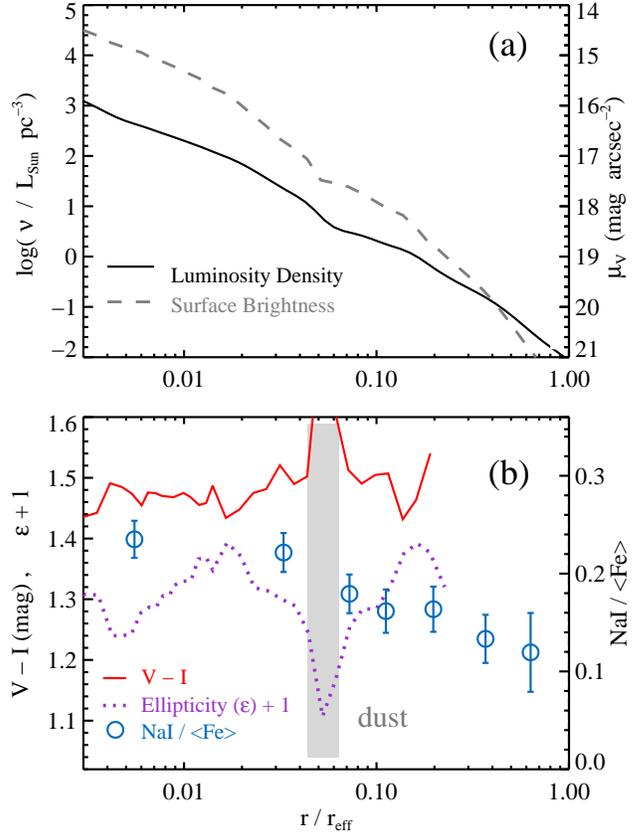,width=3.3in}
 \caption{
Major-axis photometry of NGC 2974.  (a) Surface brightness profile (dashed grey line), combining HST-based measurements by \citet{Lauer05} and ground-based measurements by \citet{Krajnovic05}.  The solid black line illustrates the corresponding deprojected luminosity density.  (b) $V$-$I$ color and ellipticity $\epsilon$ from \citet{Lauer05}, overlaid with our measurements of $\na$/$\feavg$.
}
\label{fig:phot2974}
\vspace{0.1in}
\end{figure}

Even while the causes of excess sodium, carbon, and nitrogen are murky, we have shown that the most extreme stellar populations in NGC 1023 and NGC 2974 reside in their innermost regions. 
In both objects, the strengths of $\na$, $\cat$, and CN$_1$ relative to other indices change abruptly near $0.1 \reff$, leading us to question whether this scale marks a transition in the galaxy's previous star formation environment, or a structural landmark from the assembly of distinct progenitors.  Seeking supporting evidence for unique behavior at $0.1 \reff$, we have examined \textit{Hubble Space Telescope} (HST) photometry from \citet{Lauer05} to examine each galaxy's surface brightness profile, $V-I$ color, and ellipticity out to $\approx 0.2 \reff$.  We also examined the more extended surface brightness profiles from \citet{Krajnovic05} and \citet{Scott09}, as well as our measurements of $v$ and $\sigma$ for each radial bin.
We illustrate the photometric data in Figures~\ref{fig:phot1023} and \ref{fig:phot2974}, and the kinematic data in Appendix~\ref{app:sigma} (Figure~\ref{fig:kin}).

Generally speaking, neither galaxy exhibits an abrupt feature near $0.1 \reff$ in its stellar kinematics or broadband light.  In NGC 2974, $\sigma$ is the dominant kinematic component inside $0.1 \reff$, whereas rotational $v$ dominates beyond $0.2 \reff$.  NGC 2974 also has a central dust feature, which intersects the major axis near $0.05 \reff$.  However, the absence of similar features in NGC 1023 suggests that they are not closely connected to the sharp abundance variations.

\subsection{Other Reports of IMF-Sensitive Index Gradients}
\label{sec:otherguys}

To date, few studies have sought to measure IMF gradients within individual early-type galaxies.  
Most recently, \citet{LaBarbera15b} fit radial trends in $\tio$0.89 and four temperature- and gravity-sensitive $\tio$ features in a massive early-type galaxy ($\sigma \sim 300 \kms$) at redshift $z \approx 0.057$.  Their measurements are consistent with a bimodal IMF whose slope above $\sim 0.5 \msun$ declines from a central value $\alpha \approx 3$ to a Milky Way-like IMF beyond $0.5 \reff$.  \citet{LaBarbera15b} have measured constant $\feh$ strength out to $\approx 0.2 \reff$ ($\sim 1.6$ kpc), which disfavors a varying unimodal IMF.  Our measurements of $\feh$ in NGC 1023 and NGC 2974 exhibit a similar trend at $r \leq 0.2 \reff$, and probe physical scales several times smaller than the central bin of $\sim 700$ pc adopted by \citet{LaBarbera15b}.

\citet{Zieleniewski15} have measured a rapid increase in $\na$ over the central 20 pc of M31, with no corresponding increase in $\feh$.  Similar to our interpretation above, they infer a steep gradient in [Na/Fe].  The central 40 pc of M32 do not exhibit strong radial trends in $\na$ or $\feh$.

\citet{MartinNavarro15} have measured gradients in IMF-sensitive features in three galaxies, using long-slit data covering 4500-10,000 \AA.  Near the centers of their two high-mass galaxies, NGC 4552 and NGC 5557, they find that $\na_{\rm SDSS}$ varies steeply relative to the total metallicity indicator [MgFe]$'$, similar to the trends we find above.  Their lower-mass object, NGC 4387, shows very little radial variation in $\na_{\rm SDSS}$ or in inferred stellar population properties, aside from total metallicity.  
Whereas we have carefully accounted for telluric absorption near 8190 \AA\, (see \S\ref{sec:sky} and Appendix~\ref{app:telluric}), \citet{MartinNavarro15} place less confidence in their telluric correction and exclude the $\na_{SDSS}$ index from their SPS models.  They do not measure the $\feh$ feature, and telluric emission restricts their assessment of the $\cat$ feature to the second line in the triplet.  However, they observe and model other features between 5800 \AA\, and 6400 \AA\, that we do not access with LRIS: most notably the Na D index and two $\tio$ indices.

\citet{MartinNavarro15} have reported a strong IMF gradient in NGC 4552, resolved to $\sim 0.1 \reff$ or $\sim 300$ pc.  They adopt the same bimodal IMF form as \citet{LaBarbera15b} and likewise measure a Milky Way-like IMF at $0.7 \reff$, while at the center of NGC 4552 they find an extreme slope $\alpha \approx 4$ above $0.5 \msun$.
%They measure the $\na_{\rm SDSS}$ and Na D indices and illustrate how those exhibit similar gradients in NGC 4552 despite different sensitivities to [Na/Fe], suggesting IMF-driven variations in $\na$.
They illustrate that $\na_{\rm SDSS}$ and Na D exhibit similar gradients in NGC 4552 despite different sensitivities to [Na/Fe], and assert that IMF gradients mainly drive the variations in $\na$, while Na D permits [Na/Fe] variations up to 0.25 dex.
While the combined coverage of Na D and $\na$ provides useful leverage, the modeling approach of \citet{MartinNavarro15} warrants caution: they fit selected line indices to SPS models by \citet{Vazdekis12}, after applying correction factors to adjust the measured indices from inferred [$\alpha$/Fe] to solar abundances.  The overt dependence on [$\alpha$/Fe] is particularly troubling for NGC 4552, where the [$\alpha$/Fe] gradient reported by \citet{MartinNavarro15} is much steeper than the trends typically observed for early-type galaxies \citep[e.g.,][]{SB07,Spolaor10}.

Using the the same modeling approach and bimodal IMF form, \citet{MN15b} reported a high-mass slope $\alpha \approx 4.0$ out to $\geq 1.5 \reff$ in the compact, high-$\sigma$ galaxy NGC 1277, with a mild trend toward $\alpha \approx 3.5$ in the central $0.5 \reff$ (600 pc).
For this object, \citet{MN15b} found strong gradients in $\na_{\rm SDSS}$, Na D, and metallicity, compared to relatively weak trends in $\tio_1$, $\tio_2$, and [Mg/Fe].  The results for NGC 1277 agree qualitatively with our observations of NGC 1023 and NGC 2974 on similar radial scales, and with widespread trends in [Z/H] and [$\alpha$/Fe]
\citep[e.g.,][]{Tamura00,Weijmans09,Greene13,Greene15}.
Finally, \citet{MN15c} have 
presented data from 24 galaxies in the CALIFA survey \citep{CALIFA}, binning each galaxy into several elliptical annuli.  Although they find a significant correlation between the inferred IMF slope and metallicity of each spectrum, they do not compare any measurements directly with $r$.  We therefore cannot assess whether galaxies in the CALIFA sample contain IMF or abundance gradients at the spatial scales we have probed for NGC 1023 and NGC 2974.

\citet{LaBarbera13} assessed stacked spectra from SDSS and noted that the individual galaxy spectra enclosed varying fractions of $\reff$.  They subdivided two of their stacked spectra ($\sigma = 100 \kms$ and $\sigma = 200 \kms$) into narrow bins of $r/\reff$ and found minimal variations in the strengths of $\tio$ indices and $\na_{\rm SDSS}$ from aperture sizes of $0.3 \reff$ to $1.4 \reff$.  In other words, their $\sigma = 200 \kms$ stack exhibited stronger IMF-sensitive features at all enclosed radii.  
Still, we note that any radial gradients in abundances or the IMF would be diluted by the luminosity-weighted SDSS apertures.  Our finding that the steepest gradients in NGC 1023 and NGC 2974 occur well inside $0.3 \reff$ may also bear upon the absence of gradients reported by \citet{LaBarbera13}.

Further advances in stellar template libraries and atmospheric models will yield improved predictions for the integrated-light signatures of simultaneous variations in abundances and the IMF.  In the meantime, models whereby abundance variations are restricted to [Z/H] or [$\alpha$/Fe] must be employed with caution, in light of abundance gradients for individual elements.
An interesting case study is provided by \citet{LaBarbera15b}, who attempt to rescale their observed index strengths to solar abundances, based on relations between a given index strength, [$\alpha$/Fe], and [Z/H].
Their approach is informed in part by their assessment of radial trends in [$\alpha$/Fe] and [C/Fe].
We speculate that the approach of \citet{LaBarbera15b} is better suited to the gravity-sensitive $\tio$ features they measure than to the $\na$ or $\cat$ features discussed herein, since [Ca/Fe] and [Na/Fe] are known to deviate from trends in [$\alpha$/Fe].
We look forward to further work that may support this example and identify other regimes where IMF gradients can be robustly disentangled from nuanced abundance variations.

\section{Conclusion}
\label{sec:conc}

We have used Keck/LRIS to analyze optical and near-infrared stellar absorption features along the major axis of two  early-type galaxies with central $\sigma$ of 217 $\kms$ and 247 $\kms$.  We have measured 13 line indices for species of H, C$_2$, CN, Na, Mg, Ca, $\tio$, Fe, and FeH, from spatially resolved spectra covering 3100-5560 \AA\, and 7500-10,800 \AA.  We have examined each index on scales $\sim 100$ pc ($1''$) near the center of each galaxy, and in larger bins extending to 4.0 kpc ($75''$ or $1.6\reff$) for NGC 1023 and 3.0 kpc ($30''$ or $0.8\reff$) for NGC 2974.

In both galaxies, radial declines in the $\feavg$ index and multiple indices of Mg, Ca, and C suggest an overall metallicity gradient and 
shallower gradients in [$\alpha$/Fe], [Ca/Fe], and [C/Fe].  However, the $\na$ index at 8190 \AA\, 
exhibits significantly steeper gradients, particularly at $r < 0.1 \reff$, or $r < 300$ pc.  The $\feh$ index at 9915 \AA\, mirrors the gradual decline of $\feavg$ rather than the steep decline in $\na$.  The data presented herein are among the first to track $\feh$ as a function of radius, and to demonstrate different radial trends in $\na$ and $\feh$, even while both indices are sensitive to cool dwarf stars.

We interpret the steep gradient in the $\na$ index as reflecting a rapid decline in [Na/Fe] over the central $\sim 300$ pc of each galaxy.  
On similar scales, the $\cat$ index declines relative to $\feavg$ and other Mg and Ca indices, to a degree that would require a very strong gradient in [Na/Fe] if our interpretation is to match models by \citetalias{CvD12a}.  
IMF gradients may contribute to the respective trends of $\na$, $\cat$, and $\feh$ in NGC 1023 and NGC 2974, but only if the IMF is bimodal, with a similar slope for $m < 0.4 \msun$ at all radii.
%flattens for $m < 0.4 \msun$ at all radii.
The CN$_1$ index also varies rapidly near the center of each galaxy, and a qualitative comparison to stellar population synthesis models indicates that this is due to a large-scale gradient in [C/Fe], with uniform enhancement in [N/Fe].

Our study poses a number of outstanding issues to be pursued as future work.  As we have emphasized above, the physical properties of stellar populations are highly degenerate with individual line indices, and some of the trends we have presented -- most notably the relative variations in $\na$, $\feh$, and $\feavg$ -- defy simple qualitative arguments.
Rigorous stellar population synthesis modeling will 
expand upon the qualitative comparisons we have performed herein and allow for more robust interpretations of the physical trends in NGC 1023 and NGC 2974.
Moreover, our search for radial IMF trends within two galaxies cannot fully inform claims of IMF variation over a large range of integrated galaxy properties.
To this end we have observed five additional galaxies spanning $\sigma \sim 140$-$400 \kms$, approaching the range explored by survey-driven investigations of IMF variation.
These objects will also strengthen our understanding of internal trends near $0.1 \reff$ and clarify whether this scale truly marks a transition point for star formation physics in galaxies.\\

\medskip

We thank the anonymous referee for thoroughly reviewing our manuscript and prompting substantial improvements.  We thank Daniel Perley for providing access to his pipeline for LRIS data reduction, and Pieter van Dokkum and Charlie Conroy for publishing a straightforward blueprint of their observing strategy.  
Marc Kassis, Luca Rizzi, and Hien Tran
at W. M. Keck Observatory provided essential support for conducting LRIS observations. 
NJM is supported by the Beatrice Watson Parrent Fellowship and Plaskett Fellowship.
AWM is supported by the Harlan J. Smith Fellowship.
Finally, we recognize the sacredness and cultural significance of Maunakea to indigenous Hawaiians, who have honored the land long before and indeed since the construction of modern astronomical facilities.  We hold great privilege and responsibility in using the Maunakea summit for scientific inquiry.

%\clearpage
%\vspace{0.2in}
\begin{appendix}

\section{A: Systematics in Line Index Measurements}
\label{app:sys}

Herein we derive our total systematic error $\epsilon_{\rm sys}$, which we have included above in \S\ref{sec:errs}. Figures~\ref{fig:index1023} and \ref{fig:index2974} above include all systematic terms, while Figures~\ref{fig:nacat}-\ref{fig:N1023fe} 
%\ref{fig:N2974fe} 
include all systematic terms except for the telluric absorption error in $\na$.

%TABLE - systematic errors in line indices
%
%{table*} sets table across whole page in emulateapj5
\begin{table*}[!t]
%\vspace{0.1in}
%\begin{small}
\begin{center}
\caption{Systematic Errors in Line Indices}
\label{tab:lineerrs}
%\leavevmode
\begin{tabular}[b]{lccccc}  %each c is a column, [c]entered.  For lines between columns, [c|c|c|c] etc.
\hline
Index & Error from $v_0$ & Error from $\sigma_0$ & Telluric absorption error & Error from emission lines & Adopted systematic error \\
\smallskip
& & & NGC 1023  /  NGC 2974 & NGC 1023  /  NGC 2974 & NGC 1023 / NGC 2974 \\
%(1) & (2) & (3) & (4) & (5) & (6) \\
\hline 
\\
CN$_1$ & 0.4\,\% & 0.9\,\% & --  & \, 0.9 \% /  1.4 \% &  \, 1.3 \% /  1.7 \%  \\    
Ca4227 & 1.3\,\% & 1.9\,\% & --  & --  & 2.3 \% \\   
C$_2$4668 & 0.4\,\% & 0.4\,\% & --  & $\;\;\;\;\;$ -- \, /  0.4 \% &  \, 0.6 \% /  0.7 \%  \\  
bTiO & 4.1\,\% & 1.4\,\% & --  & $\;\;\;\;\;$ -- \, /  5.2 \% &  \, 4.3 \% /  6.8 \%  \\  
H$\beta$ & 1.4\,\% & 0.2\,\% & --  & \, 6.0 \% /  6.0 \% & \, 6.2 \% /  6.2 \% \\ 
Mg b & 0.4\,\% & 0.9\,\% & --  & \, 0.7 \% /  1.0 \% &  \, 1.2 \% /  1.4 \% \\  
Fe5270 & 0.3\,\% & 1.4\,\% & --  & -- & 1.4 \% \\  
Fe5335 & 0.7\,\% & 2.7\,\% & --  & -- & 2.8 \% \\  
\\
NaI & 0.025 \AA\, & 5.5\,\% & \, 8.8-13.4\% /  1.2-3.8\% & -- & \hspace{-0.45in} $\left[\, 0.025^2 + \epsilon_{\rm tel}^2 + (0.055\,{\rm EW}_{\na})^2 \right]^{1/2}$ \,\AA \\  
& & & & & \hspace{-0.15in} $= \;\;\;$11.9-23.1\% /  7.6-11.8\%  \\  
%NaI & 0.025 \AA\, & 5.5\,\% & \, 15.1 \% /  5.9 \% & -- & \hspace{-0.45in} $\left[\, 0.025^2 + (0.055^2 + 0.151^2) \,{\rm EW}_{\na}^2 \right]^{1/2}$ \,\AA \\  
%& & & & & \hspace{-0.45in} $\left[\, 0.025^2 + (0.055^2 + 0.059^2) \,{\rm EW}_{\na}^2 \right]^{1/2}$ \,\AA \\  
%Na$_{\rm SDSS}$ & \,\% & \,\% & \,\% \\
\\
CaII & 1.1\,\% & 0.7\,\% & --  & -- & 1.3 \% \\  
MgI0.88 & 2.1\,\% & 5.6\,\% & --  & -- & 6.0 \% \\  
TiO0.89 & 0.03\,\% & 0.06\,\% & --  & -- & 0.07 \% \\  
FeH & 2.6\,\% & 2.3\,\% & --  & -- &  3.5 \% \\  
%% UNMASKED
%FeH & 2.5\,\% & 1.5\,\% & -- & 2.9 \% \\  
\hline
\end{tabular}
\end{center}
\begin{small}
Notes: In most cases the adopted systematic error is $\epsilon_{\rm sys} = (\epsilon_{v0}^2 + \epsilon_{\sigma0}^2)^{1/2}$, where $\epsilon_{v0}$ and $\epsilon_{\sigma0}$ are the error terms from $v_0$ and $\sigma_0$.  For indices impacted by telluric absorption or galaxy emission lines, the adopted systematic error is $\epsilon_{\rm sys} = (\epsilon_{v0}^2 + \epsilon_{\sigma0}^2 + \epsilon_{\rm tel}^2 +\epsilon_{\rm gas}^2)^{1/2}$, where $\epsilon_{\rm tel}$ is the error resulting from uncertainties in the atmospheric transmission spectrum, and $\epsilon_{\rm gas}$ is the error derived from different settings for emission line removal.  For $\na$, the adopted systematic error ranges from $11\%$ of the equivalent width near the galaxy center to $16\%$ of the equivalent width at large radii.
\end{small}
\vspace{0.2in}
\end{table*}

\subsection{A.1: Kinematic Corrections}
\label{app:sigma}

%FIGURE - Kinematic profiles
%
\begin{figure}[!b]
\centering
%\vspace{-0.15in}
  \epsfig{figure=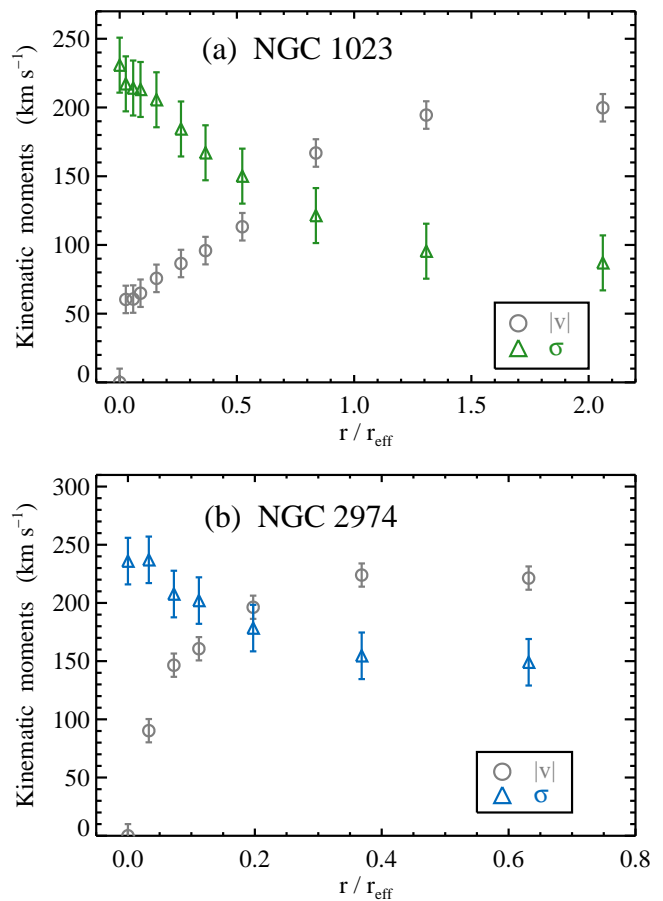,width=3.4in}
 \caption{
Radial velocity and velocity dispersion profiles for NGC 1023 and NGC 2974, as measured with {\tt pPXF}.  At each radius we have averaged $|v|$ and $\sigma$ from opposite sides of the galaxy.  We define $v =0$ at our central bin for each galaxy.  Error bars represent a constant systematic error of $10 \kms$ in $v$ and $20 \kms$ in $\sigma$.}
\label{fig:kin}
%\vspace{0.15in}
\end{figure}

%FIGURE - NaI vs. v and sigma (compare index strength and errors)
%
\begin{figure*}[!t]
\centering
\vspace{-0.2in}
  \epsfig{figure=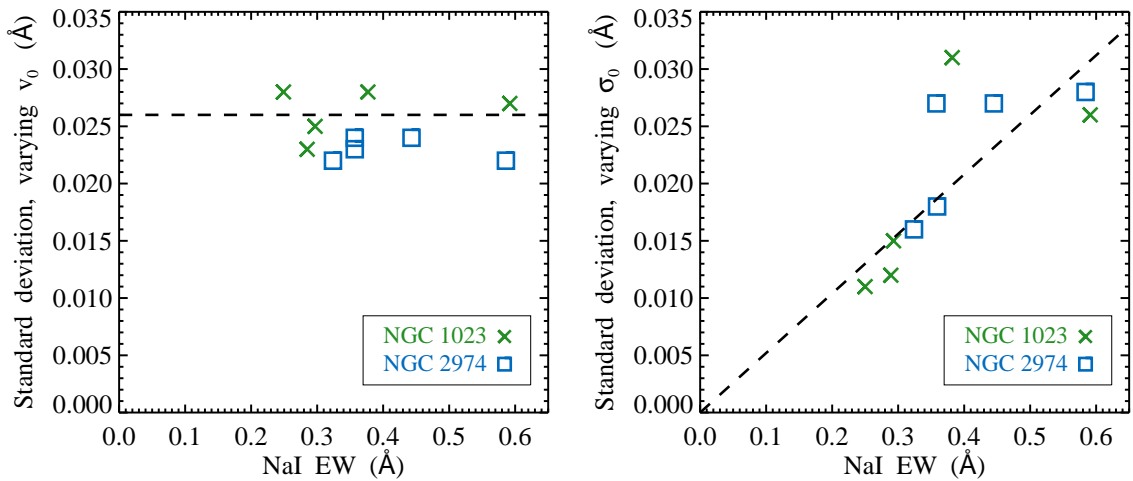,width=6.0in}
 \caption{
 \textit{Top:} Error in $\na$ feature strength for test spectra in NGC 1023 and NGC 2974, where each spectrum was shifted to rest wavelengths from a range of velocities $v_0$.  Uncertainty in the systemic velocity produces a nearly constant error over a large range of line depths.   The dashed line corresponds to an average error of 0.025 \AA, from trials sampling a conservative distribution of $v_0$.   \textit{Bottom:} error in $\na$ strength for test spectra convolved to $230 \kms$ (NGC 1023) or $245 \kms$ (NGC 2974) from a range of initial velocity dispersions $\sigma_0$.  Uncertainty in the measured velocity dispersion produces an error that increases with line depth.  The dashed line corresponds to an error of 5.5\%.
 }
\label{fig:naivsigerr}
\vspace{0.15in}
\end{figure*}

In order to make an unbiased comparison between absorption line strengths at different radii, each spectrum within a single galaxy must be shifted to the same rest frame and convolved to the same velocity dispersion.  For each spectrum, we measure $v$ and $\sigma$ using the template reconstruction procedure {\tt pPXF} from \citet{ppxf}.  We use a subset of template spectra from the MILES library of empirical stellar spectra \citep{MILES}, and perform a fit over the wavelength range 3650-5400 \AA.  We simultaneously fit for the strengths of emission lines, assuming that gas and stars exhibit the same kinematics.  Spectra on the LRIS red arm are assumed to match the kinematics derived from blue-arm data for the same spatial bins.  
We display the radial profile of $v$ and $\sigma$ for each galaxy in Figure~\ref{fig:kin}.

At high $S/N$, uncertainties in $v$ and $\sigma$ are dominated by systematic errors, particularly from template libraries that do not fully reproduce a galaxy's underlying stellar population.
For this work, we do not attempt to directly assess our systematic errors in kinematic moments, but rather test the variation in line index strengths over a conservative range of assumed $v$ and $\sigma$.   
Previous investigations using similar wavelength coverage and a variety of stellar templates have found total errors $\ltsim 10 \kms$ in $v$ and $\ltsim 20 \kms$ in $\sigma$ \citep{Barth02,ppxf}.  For the SAURON galaxy sample, \citet{Emsellem04} found scatter of 15-18 $\kms$ in comparisons between their {\tt pPXF}-based measurements of $\sigma$ and previous $\sigma$ values in the literature.
For our trials, we adopt the conservative assumption that $v$ and $\sigma$ each have errors of $20 \kms$.
We have run {\tt pPXF} using an alternative subset of MILES stellar templates and found that the resulting variations in $v$ and $\sigma$ were well within this range.

%FIGURE - NaI vs. v and sigma  (N1023 near center)
%
\begin{figure}[!b]
\centering
\vspace{0.1in}
  \epsfig{figure=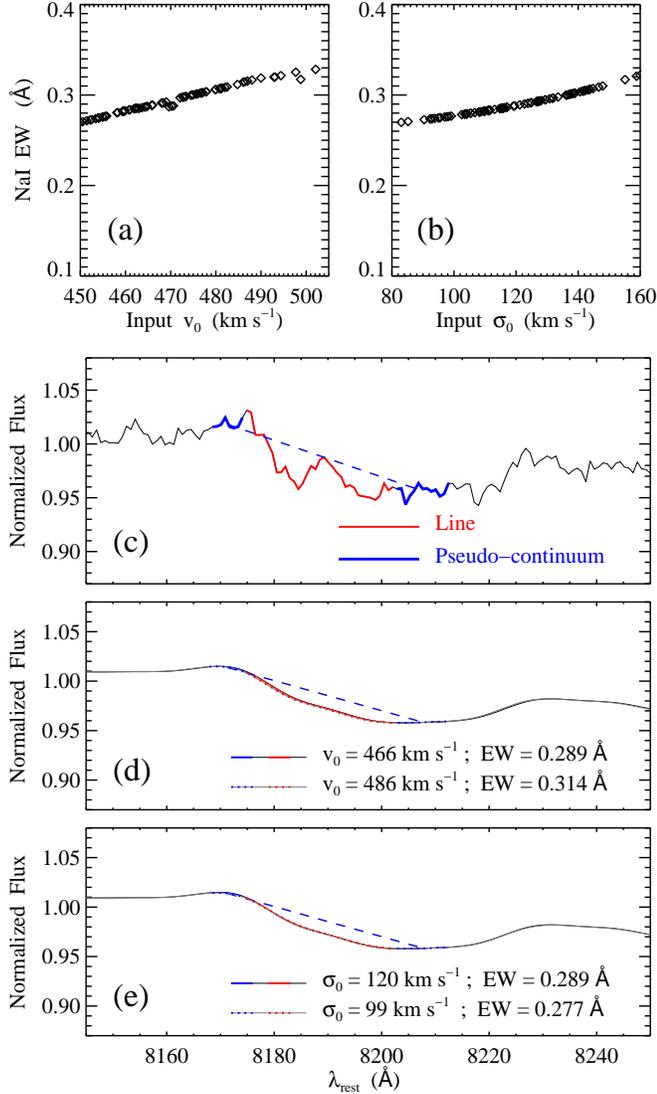,width=3.5in}
 \caption{
 (a) line strength of $\na$ feature vs. input systemic velocity ($v_0$), for a spectrum covering major-axis radii from $+0.5''$ to $+2.0''$ in NGC 1023.  (b) $\na$ line strength vs. input velocity dispersion ($\sigma_0$).  In each case, the spectrum is convolved from $\sigma_0$ to $\sigma = 230 \kms$ before measuring the line index.  (c)  rest-frame spectrum of the $\na$ feature, before convolution to $\sigma = 230 \kms$.  The red segment indicates the $\na$ line region, and the blue segments indicate the pseudo-continuum regions.   The blue dashed line is the linear fit to the pseudo-continuum, used in the index measurement.  (d) rest-frame spectrum at two separate velocity shifts.  (e) rest-frame spectrum convolved to 230 $\kms$ from two separate $\sigma_0$ values.
 }
\label{fig:naivsig}
\vspace{0.1in}
\end{figure}

%FIGURE - NaI vs. v and sigma  (N1023 large radius)
%
\begin{figure}[!b]
\centering
\vspace{0.1in}
  \epsfig{figure=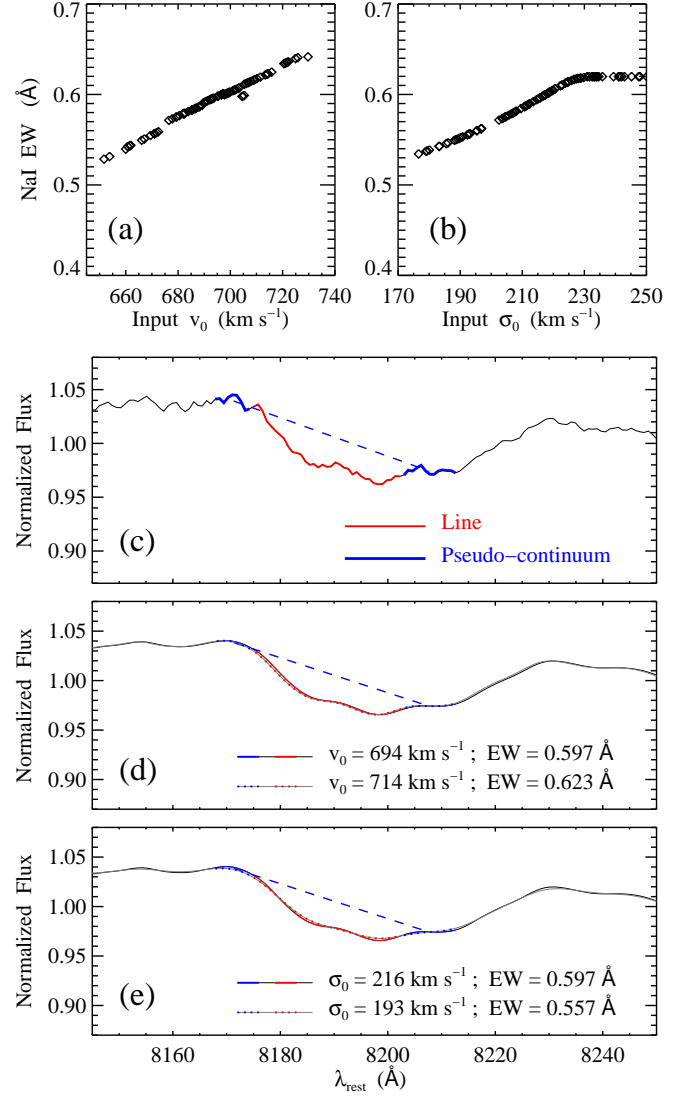,width=3.5in}
 \caption{
Panels are the same as Figure~\ref{fig:naivsig}.  This spectrum spans major-axis radii from $-50''$ to $-30''$ in NGC 1023.  The velocity dispersion is smaller than the central spectrum in 
Figure~\ref{fig:naivsig}, and the $\na$ line is significantly weaker.}
\label{fig:naivsiglow}
\vspace{1.1in}
\end{figure}

In order to assess the kinematic error terms for each line index, we have tested five spectra from NGC 1023.  The innermost spectrum spans radii from $0.5''$ to $2.0''$, with $S/N = 119$ over the wavelength range used for the {\tt pPXF} fit, and $\sigma = 216 \kms$.  The outermost spectrum spans $50''$-$75''$ and has $S/N = 38$ and $\sigma = 94 \kms$.  
For a given spectrum, we perform two sets of trials wherein we vary the assumed velocity and velocity dispersion, $v_0$ and $\sigma_0$.  In one set of trials, we fix $\sigma_0$ to the best-fit $\sigma$ from {\tt pPXF} and sample $v_0$ from a normal distribution  with a dispersion of $20 \kms$ (centered on the best-fit $v$ from {\tt pPXF}).
For each of 100 
trials sampling $v_0$, we convolve the spectrum from $\sigma_0$ to $\sigma = 230 \kms$, shift to $\lambda_{\rm rest} = \lambda_{\rm obs} / (1+v_0/c)$, and measure the absorption line indices.  For each absorption feature, we define a systematic error term $\epsilon_{v0}$ as the standard deviation in index strength over all 100 trials.  Our second set of trials assesses an analogous error term $\epsilon_{\sigma0}$ by employing a nearly identical procedure whereby $v_0$ is held fixed and $\sigma_0$ is sampled from a normal distribution with a dispersion of $20 \kms$.

In Table~\ref{tab:lineerrs} we present the values of $\epsilon_{v0}$ and $\epsilon_{\sigma0}$ for 13 different line indices, averaged over our five test spectra.  
We make particular note of the $\na$ index, which is 2-3 times stronger at the galaxy center than at large radii.  This index exhibits a similar absolute value of $\epsilon_{v0} \approx 0.025$ \AA\, at all radii, while $\epsilon_{\sigma0}$ scales approximately linearly with the index strength.  The trends in $\epsilon_{v0}$ and $\epsilon_{\sigma0}$ for both galaxies' $\na$ indices are illustrated in Figure~\ref{fig:naivsigerr}.  
Figures~\ref{fig:naivsig} and \ref{fig:naivsiglow} show individual spectra of the $\na$ region for NGC 1023, with variations in $v_0$ and $\sigma_0$ and the resulting smoothed spectra.  Although the inner spectrum (Figure~\ref{fig:naivsig}) has a much deeper $\na$ feature and much larger intrinsic $\sigma$, the change in index strength with respect to $v_0$ is similar to the outer spectrum (Figure~\ref{fig:naivsiglow}).

As indicated in Table~\ref{tab:lineerrs}, systematic errors from kinematic fitting are typically $\sim 1\%$-5\% of the line index strength.  In cases of low statistical noise, the $\epsilon_{v0}$ and/or $\epsilon_{\sigma0}$ term is comparable to the level of random variance between different data subsets.  This is true for index measurements based on high-$S/N$ spectra near the center of each galaxy, and even toward large radii for the Fe, Mg~b, and $\cat$ features.  On the other hand, the $\na$ and MgI0.88 features are relatively sensitive to the velocity shift and broadening of the underlying spectrum, such that the kinematic error terms make a substantial contribution to the overall error budget even when $S/N$ is modest.

\subsection{A.2:  Contamination from Emission Lines}
\label{app:gas}

Emission lines from warm gas are visible contaminants in our spectra throughout NGC 2974 and at the center of NGC 1023.  We have experimented with two well-known routines to remove the emission component and assess the underlying stellar absorption features: {\tt pPXF} by \citet{ppxf} and {\tt GANDALF} by \citet{gandalf}.
In addition to the stellar template library used to fit a galaxy spectrum, {\tt pPXF} supports Gaussian emission line templates at the rest wavelengths of several gas species.  The strength of each emission line is varied freely, while the emission line kinematics ($v$ and $\sigma$) are assumed to match the stellar kinematics.  Using the output template data provided by {\tt pPXF}, we separate the stellar template and emission line components, and subtract the sum of the emission lines from our original galaxy spectrum.  This is performed in the same run where we measure $v$ and $\sigma$.  

The {\tt GANDALF} routine adds more flexibility to the emission-line fitting: the gas component(s) are permitted to have different kinematics from the stars, and the relative flux and kinematics of each emission line may either be fit freely or coupled to another line.  In practice, the routine performs best when the stellar kinematics are first estimated using {\tt pPXF} (while masking the emission lines or fixing their kinematics as above), and supplied as an initial guess for the subsequent {\tt GANDALF} fit.  We have used {\tt GANDALF} successfully for NGC 2974, whose strong emission lines are easily recognized.  As before, we isolate the emission-line component of the best-fitting spectrum and subtract it from our original galaxy spectrum.
Any emission in NGC 1023 is too subtle for {\tt GANDALF} to separate cleanly, and it confuses the low-order stellar continuum with a superposition of extremely broad emission lines.  Therefore, we restrict our use of {\tt GANDALF} to NGC 2974.

We find that the best-fitting emission spectrum varies for different sets of stellar templates used in {\tt pPXF}, and with the degree to which the relative emission strengths of different species are allowed to vary in {\tt GANDALF}.  We therefore define a systematic error term $\epsilon_{\rm gas}$ as the standard deviation of trial measurements for a given absorption line index, when different settings are used for the emission line removal.  For NGC 2974, the trials include two alternative stellar template libraries -- each a subset of the empirical MILES library -- for {\tt pPXF}, and two settings for the relative emission line strengths in {\tt GANDALF}.  
In one {\tt GANDALF} trial the flux of each emission line is treated as a free parameter, and in the other trial we fix the relative strengths within the [OII], [OIII], [NI], and [ArIV] multiplets.
In NGC 2974, Balmer emission lines impact the CN$_1$, b$\tio$, and H$\beta$ absorption features.  For H$\beta$ we find that $\epsilon_{\rm gas} \approx 6\%$ of the absorption line strength, and b$\tio$ is impacted at the $\approx 5\%$ level as a result of H$\beta$ contaminating the red pseudo-continuum.  
The Mg~b and C$_2$4668 indices exhibit small systematic errors from [NI] and [ArIV] contamination, respectively.
For the average index measurements presented in \S\ref{sec:gradients} and Figures~\ref{fig:index1023}-\ref{fig:N1023fe} we have adopted {\tt pPXF} as our default tool for emission line removal, as we find it performs better at distinguishing between relatively narrow emission lines and broad variations in the galaxy continuum relative to the stellar templates.

\begin{figure}[!t]
\centering
%\vspace{0.1in}
  \epsfig{figure=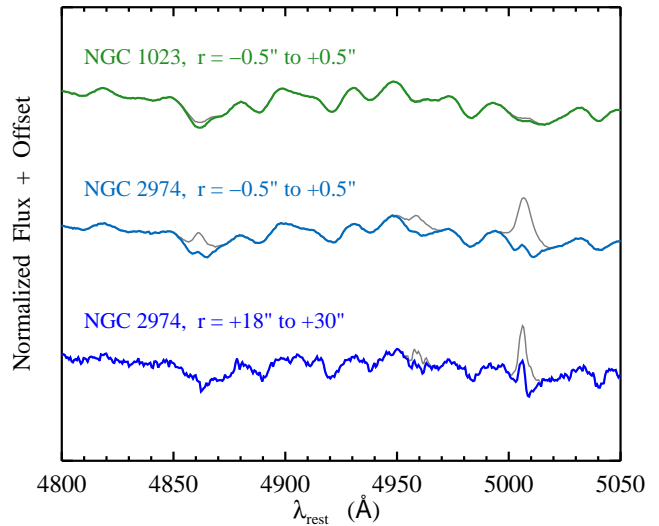,width=3.4in}
 \caption{
Galaxy spectra before and after emission line correction, zoomed in on the H$\beta$ and [OIII] features.  In each case the thin grey line represents the uncorrected spectrum, and the thick colored line represents the spectrum after correction with emission line templates in {\tt pPXF}.  From top to bottom: central bin of NGC 1023; central bin of NGC 2974; outermost bin for NGC 2974.}
\label{fig:emission}
\vspace{0.1in}
\end{figure}

For NGC 1023 we determine $\epsilon_{\rm gas}$ using two trials: a {\tt pPXF} trial with our default template library, and a trial with no emission line fitting.  A third trial with an alternative template library for {\tt pPXF} does not differ substantially for NGC 1023, and we exclude it so as not to dilute $\epsilon_{\rm gas}$.
As with NGC 2974, we find a $\approx 6\%$ impact on the H$\beta$ index, and possible low levels of contamination for the CN$_1$ and Mg~b absorption features.
In Figure~\ref{fig:emission} we show example spectra from both galaxies, before and after emission line removal.

%FIGURE - Emission line example spectra
%

\subsection{A.3:  Overlap of redshifted $\na$ and telluric H$_2$O}
\label{app:telluric}

%FIGURE - Telluric absorption trials -- example spectra
%
\begin{figure*}[!t]
\centering
%\vspace{0.1in}
  \epsfig{figure=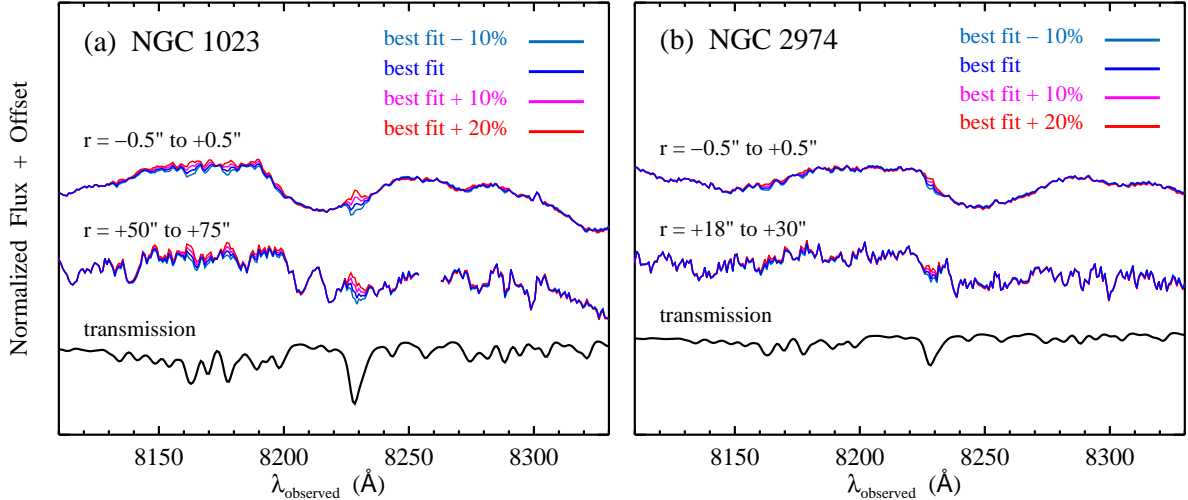,width=6.3in}
 \caption{
Galaxy spectra near the $\na$ absorption feature, after varying the strength of telluric absorption bands.  \textit{Left:} spectra from our central and outermost bins in NGC 1023.  \textit{Right:} spectra from our central and outermost bins in NGC 2974.}
\label{fig:telluric}
%\vspace{0.1in}
\end{figure*}

%FIGURE - Telluric absorption trials -- index measurements folded across the major axis
%
\begin{figure*}[!t]
%\vspace{-0.1in}
\centering
  \epsfig{figure=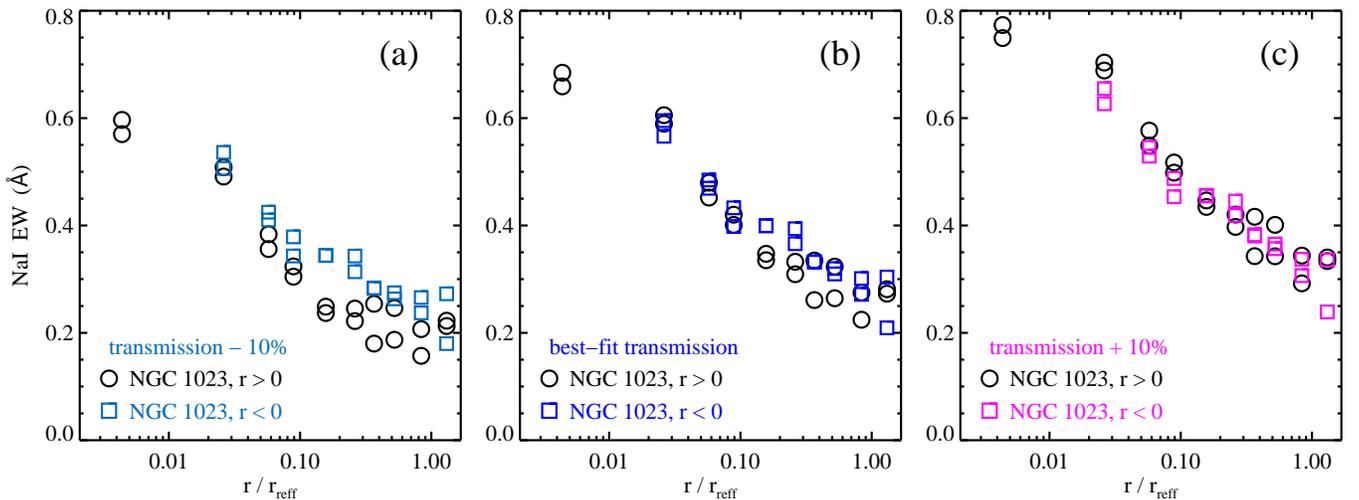,width=7.1in}
 \caption{
Comparison of $\na$ measurements in NGC 1023, for various levels of telluric correction.  The rotation curve of the galaxy will shift stellar absorption features with respect to imperfectly corrected telluric absorption bands, driving asymmetries in the line index measurements from opposite sides of the galaxy.
\textit{Left:} $\na$ measurements after weakening our best-fitting atmospheric transmission spectra by 10\%.   \textit{Middle:}  $\na$ measurements using our best-fitting transmission spectra.  \textit{Right:} $\na$ measurements after strengthening our best-fitting transmission spectra by 10\%.  The middle and right panels both represent a plausible strength for telluric absorption features in our spectra, whereas the left panel exhibits clear asymmetry between opposite sides of the galaxy as the result of poor telluric correction.}
\label{fig:telsym}
\vspace{0.2in}
\end{figure*}

We have attempted to remove telluric absorption features from our galaxy spectra by comparing them to a grid of model atmospheric transmission spectra over the observed wavelength range 9310-9370 \AA.  Although our corrected spectra look reasonable by eye, we aim to quantify the level of uncertainty near the 8190\AA\, $\na$ feature and the corresponding bias in our measurement of the $\na$ line index.  To this end, we have performed trials where we multiplied the strength of our best-fitting transmission spectrum for each science exposure by values of 0.9, 1.1, and 1.2, and repeated all data processing and analysis steps starting from telluric division of individual frames.  In Figure~\ref{fig:telluric} we display the resulting spectra for each galaxy, from our central spatial bin and an outer spatial bin.

For NGC 1023 (Figure~\ref{fig:telluric}a), contamination by the deep H$_2$O feature at 8230 \AA\, is visible for telluric correction of -10\% and +20\% relative to the best fit.  Beyond a crude visual estimate of our possible error range for telluric correction, we can take advantage of the fact that NGC 1023 and NGC 2974 both have rotational velocities $\sim 200 \kms$, such that the $\na$ feature is shifted by $\sim 10$ \AA\, over the length of our slit.  If the $\na$ line or pseudo-continuum regions are severely contaminated by a telluric feature, we should see systematic differences in our index measurements from opposite sides of the galaxy.  This is illustrated in Figure~\ref{fig:telsym}.  For each panel, we have overplotted all four subsamples used to compute our random errors in the $\na$ index (see \S\ref{sec:errs}).  For the trial where we adjusted our transmission spectra by -10\% (Figure~\ref{fig:telsym}a), our measurements of the $\na$ index on opposite sides of the galaxy are clearly offset.  We see similar asymmetry for the trial with +20\% adjustment, though it is not displayed in Figure~\ref{fig:telsym}.  Our other two trials (no adjustment and +10\%, corresponding to Figures~\ref{fig:telsym}b and \ref{fig:telsym}c) both exhibit overlapping measurements from opposite sides of the galaxy and do not produce obviously poor spectra.  We therefore estimate our plausible systematic error from telluric correction in terms of the difference between our $\na$ index measurements from these latter two trials.  
Specifically, we have adjusted our final $\na$ index measurement in each spatial bin of NGC 1023 to the average value from the two trials.  In each bin we define the 1-$\sigma$ systematic error term $\epsilon_{\rm tel}$ as the percent deviation between the two trials, equal to 
%$1/\sqrt2$ 
$\frac{1}{\sqrt2}$
times the percent difference in index measurements.  

Our random errors in each line index are derived from subsets of data spanning both sides of the galaxy (\S\ref{sec:errs}), and therefore some of the scatter in Figure~\ref{fig:telsym} is already incorporated in the random error term for $\na$.  However, Figure~\ref{fig:telsym} also exhibits a systematic offset between the three panels, indicating that poor telluric correction introduces an overall bias in addition to increased scatter from the shift in observed wavelengths of $\na$.   
We therefore keep the $\epsilon_{\rm tel}$ term as defined above, to account for the plausible bias level from over- or under-corrected telluric features.

%FIGURE - Comparison of NaI and NaI_SDSS spectral regions
%
\begin{figure}[!b]
\centering
\vspace{0.2in}
  \epsfig{figure=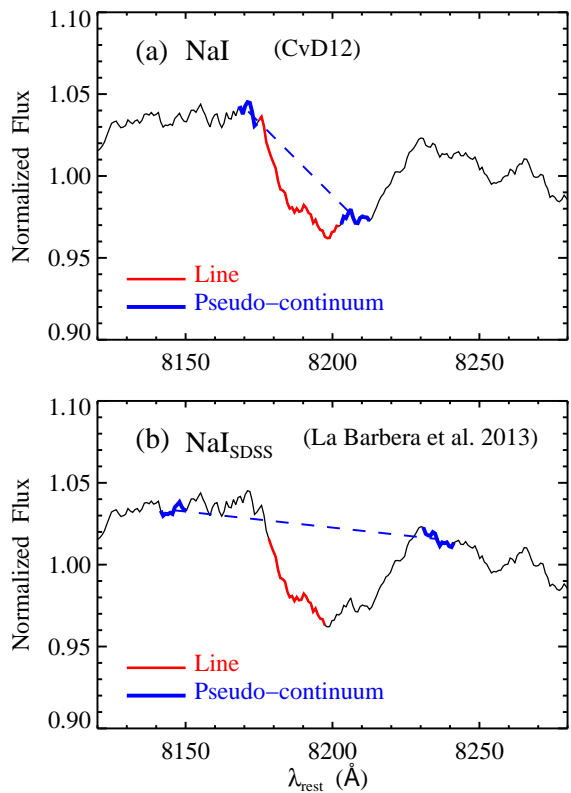,width=3.0in}
 \caption{Comparison of the $\na$ and $\na_{\rm SDSS}$ indices defined in Table~\ref{tab:linedefs}.  In each panel the red segment indicates the line region, and the blue segments indicate the pseudo-continuum regions.   The blue dashed line is the linear fit to the pseudo-continuum, used in the index measurement.  The example galaxy spectrum is the same as in Figure~\ref{fig:naivsig}, prior to convolution to $\sigma = 230 \kms$.  
 }
\label{fig:nasdss}
%\vspace{0.15in}
\end{figure}
%

%FIGURE - Comparison of NaI and NaI_SDSS index strength vs. <Fe>
%
\begin{figure}[!b]
\vspace{0.1in}
\centering
  \epsfig{figure=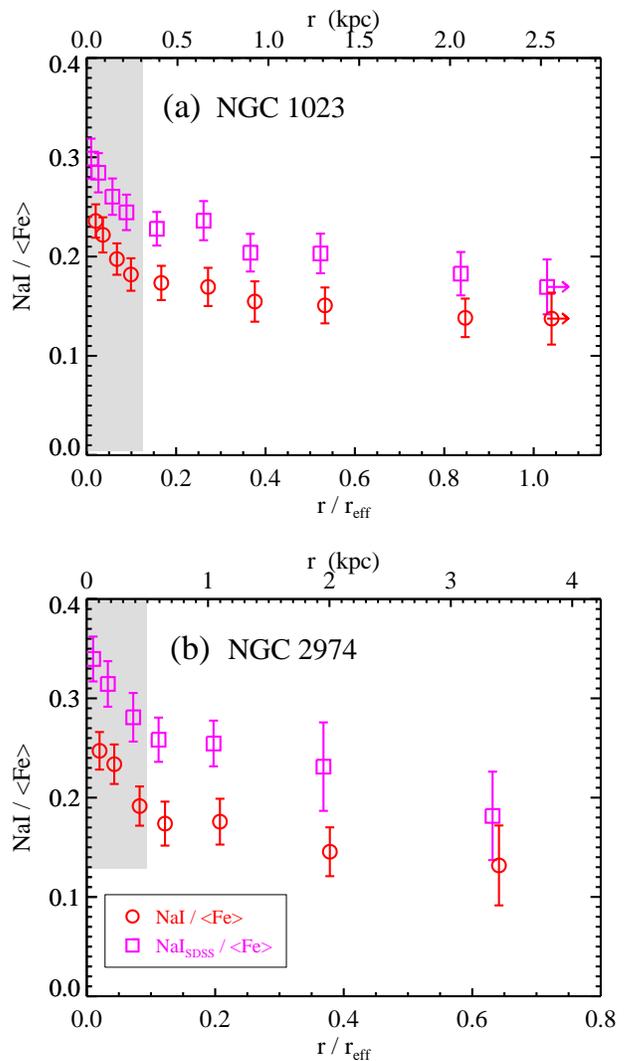,width=3.2in}
\caption{Ratio of $\na$ and $\na_{\rm SDSS}$ to $\feavg$ as a function of radius.  \textit{Left}:  NGC 1023, with the outermost spatial bin extending to $1.6 \reff$.  \textit{Right:} NGC 2974, with the outermost spatial bin extending to $0.8\reff$.  The grey shaded area in each panel indicates the scale of the single-aperture measurement by \citet{vDC12}.
 }
%\vspace{0.1in}
\label{fig:nasdssr}
%\vspace{0.15in}
\end{figure}

We observed NGC 2974 during the second half of our observing night, and telluric H$_2$O in our spectra is not as strong as for NGC 1023 (see Figure~\ref{fig:telluric}b).  Yet our -10\% and +20\% atmospheric transmission trials once again produce offsets in the $\na$ index measurements from opposite sides of the galaxy.  Therefore we define $\epsilon_{\rm tel}$ for NGC 2974 in the same manner as for NGC 1023, and adjust our average index values accordingly.
For NGC 2974 we compute a maximum $\epsilon_{\rm tel}$ of $3.8\%$, versus $13.4\%$ for NGC 1023.
For both galaxies we add $\epsilon_{\rm tel}$ in quadrature with other systematic terms, as indicated in Table~\ref{tab:lineerrs}.

%\clearpage
%\vspace{0.2in}

\section{B: Alternative Definitions of the Sodium 8190\AA\, and FeH Wing-Ford Indices}
\label{app:altna}

%FIGURE - Comparison of different FeH index definitions
%
\begin{figure*}[!t]
\centering
  \epsfig{figure=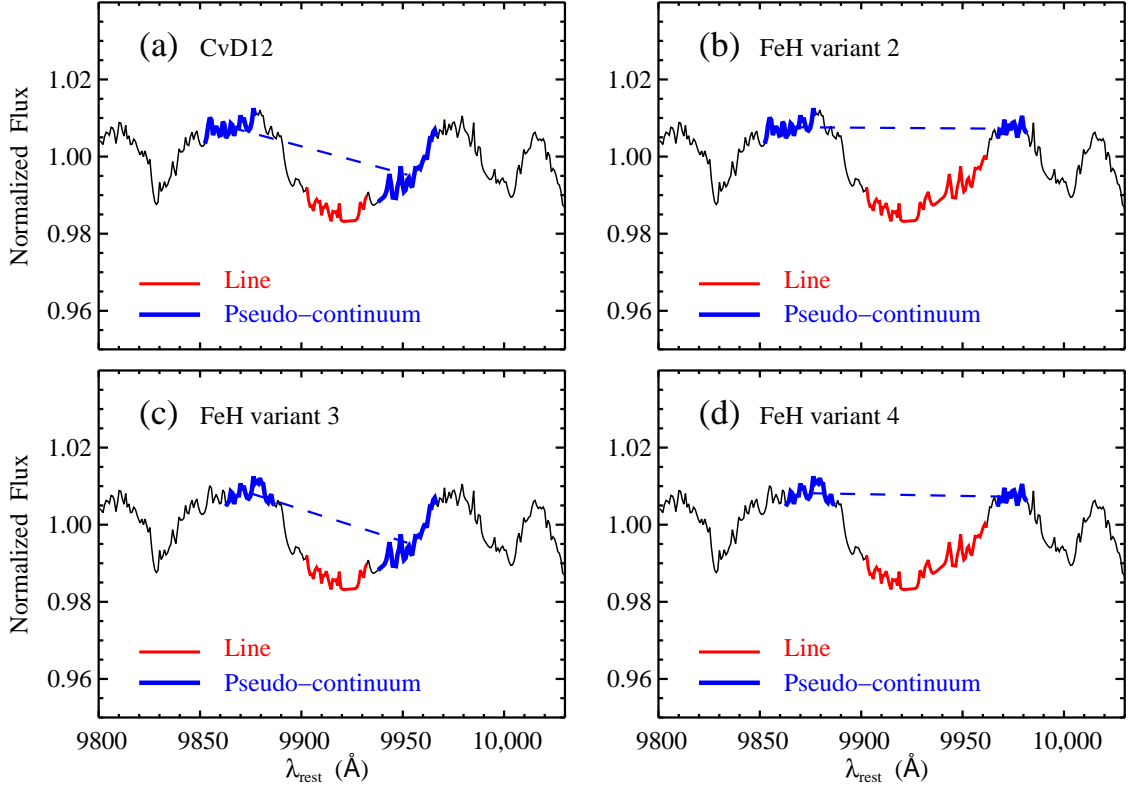,width=6.0in}
 \caption{Comparison of the $\feh$ index definition from \citetalias{CvD12a} and three variants explored herein.   In each panel the red segment indicates the line region, and the blue segments indicate the pseudo-continuum regions.   The blue dashed line is the linear fit to the pseudo-continuum, used in the index measurement.  The example galaxy spectrum is the innermost bin ($-5''$ to $+5''$) for NGC 1023, with no velocity convolution.  For visualization purposes, the spectrum has been divided by a fourth-order polynomial approximating the continuum from 9800-10,050 \AA.
 }
\label{fig:fehvar}
%\vspace{0.3in}
\end{figure*}

%
%FIGURE - Comparison of NaI and NaI_SDSS index strength vs. r
%
\begin{figure*}[!t]
\vspace{-0.2in}
\centering
  \epsfig{figure=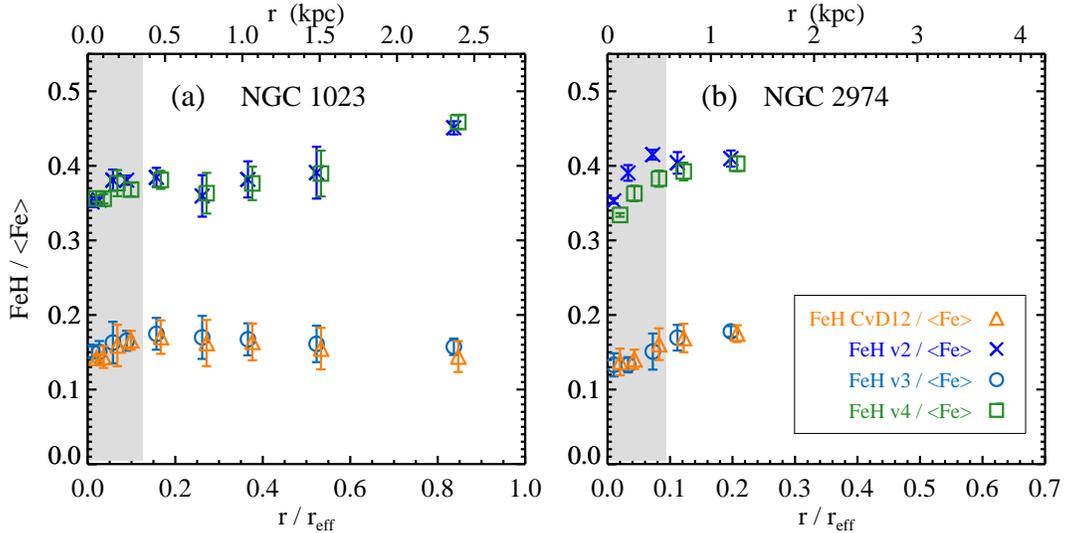,width=5.8in}
 \caption{Ratio of $\feh$ to $\feavg$ as a function of radius, using the four definitions of the $\feh$ index in Figure~\ref{fig:fehvar}.   \textit{Left}:  NGC 1023, with good data for $\feh$ extending to $1.0 \reff$.  \textit{Right:} NGC 2974, with good data for $\feh$ extending to $0.26\reff$.  The grey shaded area in each panel indicates the scale of the single-aperture measurement by \citet{vDC12}.
 }
\label{fig:fehvarr}
\vspace{0.25in}
\end{figure*}

In recent investigations of IMF-sensitive absorption features, \citetalias{CvD12a} and \citet{LaBarbera13} have proposed two different definitions of the $\na$ absorption index near 8190 \AA.  We compare the corresponding line and pseudo-continuum regions in Figure~\ref{fig:nasdss}.  In the version by \citetalias{CvD12a}, both pseudo-continuum regions are immediately adjacent to the line region, such that the red pseudo-continuum lies directly on the overlapping $\tio$ band.  The $\na_{\rm SDSS}$ index from \citet{LaBarbera13} defines a narrower line region and moves both pseudo-continua to regions outside the $\na$+$\tio$ blend.  This generates a shallower linear fit to the continuum and produces larger equivalent widths.  However, we find that the offset between our measured $\na$ and $\na_{\rm SDSS}$ line strengths is nearly the same at all radii in NGC 1023 and NGC 2974.  In Figure~\ref{fig:nasdssr} we show the ratios $\na$/$\feavg$ and $\na_{\rm SDSS}$/$\feavg$ as a function of radius in NGC 1023 and NGC 2974.  For both definitions, the relative strength of the sodium index rises steeply from $0.1\reff$ to the galaxy center.  For all figures and discussion above, we follow the $\na$ index definition from \citetalias{CvD12a}.

The relative strengths of the $\feh$, $\na$, and $\feavg$ absorption features are crucial for our interpretation of abundance versus IMF gradients in NGC 1023 and NGC 2974.  The $\feh$ index is notoriously difficult to measure, as spectra near 9900 \AA\, suffer from bright telluric emission lines and relatively low instrumental throughput.  While the Wing-Ford feature corresponds to the principal FeH bandhead in this region, the shape of the nearby pseudo-continuum is also influenced by overlapping bands of TiI, $\tio$, and CrH \citep[e.g.,][CvD12]{McLean03,Cushing05}.   
To check whether our observed radial trends in $\feh$ are sensitive to the precise placement of the line and pseudo-continuum regions, we have repeated our measurements using four variants of the index definition.  These are illustrated in Figure~\ref{fig:fehvar}.  
The first variant is the definition by \citetalias{CvD12a}, which we list in Table~\ref{tab:linedefs} and use for all analysis and figures above.  In this version, the line region is defined to match the deepest part of the FeH bandhead, possibly reducing contamination from overlapping $\tio$.  Our second variant 
%(Figure~\ref{fig:fehvar}b) 
instead uses a larger portion of the FeH trough (9902.3-9962.3 \AA) and moves the red pseudo-continuum accordingly to 9967.3-9982.3 \AA.  The third and fourth variants 
%(Figures~\ref{fig:fehvar}c,d) 
match the first two, except with the blue pseudo-continuum moved from 9852.3-9877.3 \AA\, to 9862.3-9887.3 \AA.

We display the radial trends of $\feh$/$\feavg$ for each variant in Figure~\ref{fig:fehvarr}.  None of the adopted $\feh$ variants alter our main finding: that $\feh$ decreases relative to $\feavg$ going from $0.1 \reff$ into the center of each galaxy, and thereby opposes the trend in $\na$/$\feavg$ at these same radii.  Yet there is evidence of some deviation at $r > 0.5 \reff$ in NGC 1023, such that the variants using the full absorption trough (variants 2 and 4) show an increase in $\feh$ toward large $r$, whereas the original definition by \citetalias{CvD12a} and our similar variant 3 show a slight decrease toward large $r$.
This may be caused by a faint sky line near 9970 \AA\, rest, which overlaps with the red pseudo-continuum in our full-trough variants.  Traces of this line are visible in Figure~\ref{fig:skysub}b and for the two outermost spectra in Figure~\ref{fig:specrad}c.

\end{appendix}

%\clearpage 
\vspace{0.1in}

%% Papers we'll need to add somewhere above (already in master.bib)
%  LaBarbera15b
%  Zieleniewski15

\nocite{CvD12a}
\nocite{Spiniello14}
\nocite{LaBarbera13}
\nocite{Cenarro01}
%\nocite{Serven05}
\nocite{Trager98}
\nocite{Trager00b}
\nocite{Greene15}
%\nocite{skycalcJones}
\nocite{Meynet02a}
\nocite{Meynet02b}
\nocite{Chiappini05}
\nocite{Hirschi07}
\nocite{Cohen05}
\nocite{Renzini08}
\nocite{Fenner04}
\nocite{Conroy12}
\bibliographystyle{apj}
\bibliography{master}

\end{document}